\documentclass[amsmath,amssymb,prb,twocolumn]{revtex4}
\usepackage[dvips]{graphicx}
\usepackage{bm}% bold math

\usepackage{color}
\newcommand{\YU}[1]{#1}

\begin{document}
\title{
Full-counting statistics of information content and heat quantity in the steady state and the optimum capacity
}
\author{Yasuhiro Utsumi}
\address{Department of Physics Engineering, Faculty of Engineering, Mie University, Tsu, Mie 514-8507, Japan}

\begin{abstract}
We consider a bipartite quantum conductor and analyze fluctuations of heat quantity in a subsystem as well as self-information associated with the reduced-density matrix of the subsystem. 
By exploiting the multi-contour Keldysh technique, we calculate the R\'enyi entropy, or the information generating function, subjected to the constraint of the {\it local} heat quantity of the subsystem, from which the probability distribution of conditional self-information is derived. 
We present an equality that relates the optimum capacity of information transmission and the R\'enyi entropy of order 0, which is the number of integer partitions into distinct parts. 
We apply our formalism to a two-terminal quantum dot. 
We point out that in the steady state, the reduced-density matrix and the operator of the {\it local} heat quantity of the subsystem may be commutative.
\end{abstract}

\date{\today}

\maketitle

\newcommand{\mat}[1]{\mbox{\boldmath$#1$}}

\section{Introduction}
\label{sec:intro}

\begin{figure}[ht]
\includegraphics[width=0.7 \columnwidth]{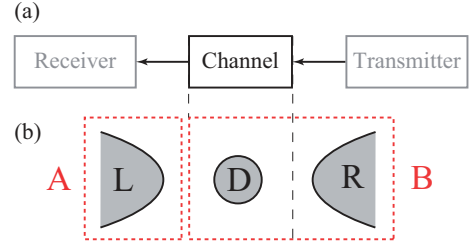}
\caption{
(a) A model of the communication system. 
A transmitter-generated signal is sent through a channel to a receiver. 
We focus on the signal transmission process through the channel. 
(b) A quantum conductor (single-level quantum dot) coupled to the left and right leads. 
We regard the quantum conductor as the communication channel. 
The right lead corresponds to the transmitter, which generates thermal and shot noise as signals. 
The left lead corresponds to the receiver side. 
The electron temperature of the left lead is set to zero in order to suppress the intrinsic thermal noise in the receiver side. 
We regard the left lead as subsystem $A$ and the quantum conductor and the right lead as subsystem $B$. 
}
\label{setup}
\end{figure}

The laws of physics limit the performance of information processing~\cite{Bremermann1967,Landauer1973,Landauer1996,Lloyd2000}. 
The quantum limits of information transmission through a quantum communication channel have long been discussed~\cite{Bremermann1967,Lloyd2000,Lebedev1963,Gordon1964,Takahashi1965,Pendry1983,CavesRMP1994,BlencowePRA2000}. 
In information theory, a model communication system consists of a transmitter, a channel, and a receiver~\cite{Shannon1948} [Fig.~\ref{setup}~(a)]. 
The physically relevant part is the channel through which a signal produced by the transmitter reaches the receiver. 
A measure of the performance of a channel is capacity $C$, the maximum possible rate at which information can be transmitted without error. 
More precisely, let $I$ be the amount of information content transmitted during a given measurement time $\tau$. 
Then the rate of information transmission always satisfies $I/\tau \leq C$. 
The capacity of a wideband quantum channel for a given average signal power $P$ is~\cite{Lebedev1963,Gordon1964,Pendry1983,CavesRMP1994,BlencowePRA2000} (we set $\hbar=k_{\rm B}=e=1$), 
%------------------------------------------------------------------------------
\begin{align}
C_{\rm WB}(P) = \sqrt{ \frac{\pi}{3} N_{\rm ch} P } \, , \label{eqn:cwb}
\end{align}
%------------------------------------------------------------------------------
where $N_{\rm ch}$ is the number of channels. 
For a fermionic channel~\cite{Pendry1983,BlencowePRA2000}, when the information is carried by electrons, $N_{\rm ch}=1/2$. 
For a bosonic channel~\cite{Lebedev1963,Takahashi1965,CavesRMP1994,BlencowePRA2000} and for a fermionic channel with electrons and holes, $N_{\rm ch}=1$. 

\YU{
The square root dependence on $P$ of Eq.~(\ref{eqn:cwb}) can be deduced from the energy-time uncertainty relation~\cite{Bremermann1967,Pendry1983,Lloyd2000}. 
Here, we briefly estimate the capacity following Ref.~\onlinecite{Pendry1983}. 
Roughly speaking, it is not possible to distinguish energy quanta smaller than $\delta E \sim \hbar/(2 \tau)$~\cite{Bremermann1967,Pendry1983}
(see also Ref.~\onlinecite{Lloyd2000}). 
Suppose one bit of information content is conveyed by the arrival or non-arrival of an electron. 
Since there are $N_{\rm ch}$ channels, an energy window larger than $I \delta E/N_{\rm ch}$ is needed in order to send $I$ bits of information content.
This energy window is accompanied by the energy current, i.e., the signal power, which is estimated by using the Landauer formula for heat current~\cite{Sivan1986} as, $P=E/\tau \geq N_{\rm ch} h^{-1} \int_0^{I \delta E/N_{\rm ch}} E' dE' = (I  \delta E)^2/(2h N_{\rm ch})$. 
By rewriting this inequality as $I \leq \sqrt{2 h N_{\rm ch} P}/\delta E$, and by replacing $\delta E$ with $\hbar/(2 \tau)$, we obtain $I/\tau \leq 4 \sqrt{\pi N_{\rm ch} P}$, which is consistent with Eq.~(\ref{eqn:cwb}). 
}

In information theory, the channel capacity is defined as the mutual information per second between input signal and output signal maximized with respect to the distribution of the input signal~\cite{Shannon1948}. 
Equation~(\ref{eqn:cwb}) is indeed the {\it optimum capacity} $C_{\rm opt}$, which is the capacity further maximized with respect to input states and output measurement schemes~\cite{CavesRMP1994}. 
The optimum capacity is the logarithm of the size of the Fock subspace containing electrons with total energy $E= P \tau$. 
It turned out that the optimum capacity is the {\it partition function} of the theory of partition~\cite{Andrews2004}, i.e., the number of ways to write a positive integer as the sum of positive integers that satisfy a certain condition depending on the statistics of particles~\cite{CavesRMP1994,BlencowePRA2000}. 

%----------------------------------------------------------
%----------------------------------------------------------

In the present paper, we discuss information transmission through a mesoscopic quantum electric conductor connected to a left lead and a right lead. 
We regard the right lead as the transmitter generating thermal and shot noise as signals and regard the quantum conductor as the channel. 
The left lead corresponds to the receiver side, see Fig.~\ref{setup} (b). 
Temperatures and chemical potentials of the receiver side and the transmitter side can be different. 
\YU{We will set the energy origin at the chemical potential.}
Therefore, the signal power $P$ would be the heat current $Q/\tau$ rather than the energy current $E/\tau$. 

In the previous theories~\cite{Gordon1964,Lebedev1963,Pendry1983,CavesRMP1994,BlencowePRA2000}, an ideal quantum channel was considered. 
For mesoscopic quantum electric conductors, the scattering theory was developed to analyze the entropy current~\cite{Sivan1986} as well as the capacity~\cite{Akkermans2009}. 
However, there are not many works in this direction. 
In the present paper, we analyze the information content obtained by the receiver side. 
For this purpose, we bipartition the whole system into the receiver side (subsystem $A$) and the transmitter side (subsystem $B$), which includes the channel [Fig.~\ref{setup} (b)]. 
Subsystem $A$ consists of the left lead. 
Subsystem $B$ consists of the right lead and the quantum conductor. 
We introduce a reduced-density matrix of subsystem $A$ by tracing out subsystem $B$ degrees of freedom, 
%------------------------------------------------------------------------------
$\hat{\rho}_A = {\rm Tr}_B \hat{\rho}$.
%------------------------------------------------------------------------------
Then we perform a projective measurement of ``{\it local} heat quantity" of subsystem $A$, $Q_A$, \YU{or its dimensionless equivalent}, $S_A=\beta_A Q_A$, where $\beta_A$ is the inverse temperature of subsystem $A$. 
The reduced-density matrix after obtaining outcome $S_A$ is, 
%------------------------------------------------------------------------------
\begin{align}
\hat{\rho}_{A,S_A}  = \frac{ \hat{\Pi}_{S_A} \hat{\rho}_A \hat{\Pi}_{S_A} }{P(S_A)} \, , \;\;\;\
P(S_A) = {\rm Tr}_A \left( \hat{\Pi}_{S_A} \hat{\rho}_A \right) \, , 
\end{align}
%------------------------------------------------------------------------------
where $\hat{\Pi}_{S_A}$ is a projection operator and $P (S_A) $ is the probability of obtaining the measurement outcome $S_A$. 
The operator of the conditional self-information~\cite{Cover2006} associated with the state of electrons with signal power $P_A=Q_A/\tau$ would be $\hat{J} = - \ln \hat{\rho}_{A,S_A}$ (hereafter we choose base $e$). 
The operator $\hat{J}$ is formally the `entanglement Hamiltonian'~\cite{LiPRL2008,FrancisSong2012,Petrescu2014} subjected to the ``local heat quantity" constraint. 
The purpose of this paper is to analyze the probability distribution of the conditional self-information beyond its average value, i.e. the conditional entropy. 
\YU{
In fact, the conditional self-information is a random variable, and thus one can consider its probability distribution function; 
%------------------------------------------------------------------------------
\begin{align}
P_{S_A}(J) &= {\rm Tr}_A \left[ \hat{\rho}_{A,S_A} \delta ( J + \ln \hat{\rho}_{A,S_A} ) \right] \label{condpdf}
\, . 
\end{align}
%------------------------------------------------------------------------------
By exploiting the orthonormal decomposition of the density matrix $\hat{\rho}_{A,S_A}=\sum_n p_n |n \rangle \langle n|$, where $|n \rangle$ is an orthonormal set and $p_n$ are eigenvalues of $\hat{\rho}_{A,S_A}$, 
the probability distribution function is written as 
$P_{S_A}(J) = \sum_n p_n \delta ( J + \ln p_n )$
 (see e.g., Chapter 2.7 in Ref.~\onlinecite{Fano1961}). 
It is convenient to introduce the characteristic function or the information-generating function~\cite{Golomb1966,Guiasu1985}, the Fourier transform of the probability distribution function, 
$ \int dJ e^{i \xi J} P_{S_A}(J) = \sum_n p_n^{1-i \xi}$, 
which may be regarded as the R\'enyi entropy of order $\alpha=1-i \xi$~\cite{Renyi1960,note1}. 
As we will see later in Eq.~(\ref{c_renyi}), the Fourier transform of the probability distribution function (\ref{condpdf}) is related to the R\'enyi entropy of order $M$, 
}
%------------------------------------------------------------------------------
\begin{align}
S_{M}(S_A) = {\rm Tr}_A \left[ \left( \hat{\Pi}_{S_A} \hat{\rho}_A \hat{\Pi}_{S_A} \right)^M \right] \, . 
\label{renyi_q}
\end{align}
%------------------------------------------------------------------------------
%
The main message of the present paper is that, when the thermal noise of the receiver side is suppressed $\beta_A \to \infty$, there exists a universal relation similar to Jarzynski equality~\cite{Jarzynski1997,Campisi2011}, which connects the probability distribution of the conditional self-information, the R\'enyi entropy of order 0, and the optimum capacity, 
%------------------------------------------------------------------------------
\begin{align}
\left \langle e^J \right \rangle_{Q_A} = S_0(Q_A) \approx \exp \left(\tau C_{\rm opt}(P_A) \right) \, .  \label{jareq}
\end{align}
%------------------------------------------------------------------------------
We demonstrate that in our case, Eq.~(\ref{jareq}) is the partition function of integer partitions into distinct parts. 

Here we note a subtle issue concerning the definition of the operator of the ``{\it local} heat quantity" of subsystem $A$. 
In general, the reduced-density matrix is not diagonal in the eigenbasis of the operator of ``{\it local} heat quantity" $ [ \hat{\rho}_A , \hat{Q}_A ] \neq 0 $ (see Eq.~(\ref{eqn:heat_op}) for the definition of $\hat{Q}_A$). 
%, Eq.~(\ref{eqn:noncommutativity}). 
It is a manifestation of the noncommutativity between the Hamiltonian of the subsystem $A$, $\hat{H}_A$, and the full Hamiltonian, 
%------------------------------------------------------------------------------
\begin{align}
\hat{H}=\hat{H}_A + \hat{H}_B + \hat{V} \, , 
\end{align}
%------------------------------------------------------------------------------
which includes the coupling between the two subsystems $\hat{V}$. 
This noncommutativity causes difficulties in constructing the thermodynamics of an open quantum system coupled strongly to reservoirs~\cite{Ludovico2014,EspositoPRL2015,EspositoPRB2015,OchoaPRB2016,Ludovico2016}. 
In our case, it causes difficulties in dealing with the projection operator in Eq.~(\ref{renyi_q}). 
In the present paper, we concentrate on the steady state, where the time translational invariance is restored and the coupling energy is neglected as compared with the net energy transfer between the subsystems. 
In such a case, we can regard $ [ \hat{\rho}_A , \hat{Q}_A ] \approx 0 $ and circumvent this problem. 

\YU{
Another purpose of the present paper is to extend the multi-contour Keldysh Green function technique~\cite{Ansari2015,Ansari2015b,YU2017,Nazarov2011,YU2015,YU2018fqmt}. 
From this point of view, the present paper relies on our previous works~\cite{YU2017,YU2015,YU2018fqmt}. 
In Ref.~\onlinecite{YU2015}, we developed the replica trick to calculate the R\'enyi entropy in the non-equilibrium steady state. 
In Ref.~\onlinecite{YU2017}, we accounted for the local particle number constraint to analyze the accessible entanglement. 
In the present paper, we will account for the local heat quantity constraint, Eqs.~(\ref{renyi_constraint}) and (\ref{eqn:renyi_multi}). 
In this way, we are able to calculate the information channel capacity subjected to signal power constraint, which connects thermodynamics and communication theory. 
A celebrated paper by Shannon~\cite{Shannon1948} demonstrated that the channel capacity of the Gaussian channel depends on the bandwidth $B$, the average signal power $P$ and the noise power $P_{\rm Noise} $ as $C=B \ln (1+P/P_{\rm Noise}) $. 
In the present paper, we discuss the quantum version of the channel capacity. 
The flows of R\'enyi entropy and energy have been discussed also in Ref.~\onlinecite{Ansari2015b}. 
}

The structure of the paper is as follows. 
In Sec.~\ref{sec:igf}, we introduce probability distributions and information-generating functions. 
Then we present a universal relation, Eq.~(\ref{jareq1}). 
In Sec.~\ref{sec:mcKeldysh}, we summarize the multi-contour Keldysh generating function~\cite{Ansari2015,Ansari2015b,YU2017,Nazarov2011,YU2015,YU2018fqmt}. 
In Sec.~\ref{sec:srlm}, we apply our formalism to a resonant-level model, and then in Sec.~\ref{sec:opt_cap} we derive the optimum capacity. 
In Sec.~\ref{sec:pdf}, we focus on energy-independent transmission cases. 
In Sec.~\ref{sec:rtcb}, we turn to the resonant tunneling condition. 
We also discuss the commutability of the reduced-density matrix and the operator of the ``local heat quantity" in the presence of the Coulomb interaction in Sec.~\ref{sec:couint}. 
In Sec.~\ref{sec:previous_theories}, we discuss differences between our approach and the previous quantum information theory approach~\cite{CavesRMP1994,BlencowePRA2000}. 
In Sec.~\ref{sec:summary}, we summarize our findings.

\section{Information-generating function}
\label{sec:igf}

\subsection{Joint probability distribution and conditional probability distribution}

We assume that initially the two subsystems $A$ and $B$ are decoupled. 
Each subsystem is in equilibrium: 
%------------------------------------------------------------------------------
\begin{align}
\hat{\rho}_{A(B) {\rm eq}} = e^{- \beta_{A(B)} (\hat{H}_{A(B)}-\mu_{A(B)} \hat{N}_{A(B)})}/Z_{A(B) {\rm eq}} \, , 
\label{eqn:Z_eq}
\end{align}
%------------------------------------------------------------------------------
where $\beta_{A(B)}$ and $\mu_{A(B)}$ are the inverse temperature and the chemical potential of subsystem $A(B)$, respectively. 
The equilibrium partition function $Z_{A(B) {\rm eq}}$ ensures the normalization condition ${\rm Tr}_{A(B)} \hat{\rho}_{A(B) {\rm eq} }=1$. 
\YU{Explicitly, the initial density matrix is $\hat{\rho}(t<0)=\hat{\rho}_{A {\rm eq}} \, \hat{\rho}_{B {\rm eq}}$. 
At $t=0$, we switch on the coupling $\hat{V}$ and let the total system evolve untill $t=\tau$. 
Then we trace out the subsystem $B$ and obtain the reduced density matrix of the subsystem $A$ as, 
%------------------------------------------------------------------------------
\begin{align}
\hat{\rho}_A (\tau) = {\rm Tr}_B \hat{\rho} (\tau) \, , 
\;\;\;\
\hat{\rho} (\tau) = e^{-i \hat{H} \tau} \hat{\rho}_{A {\rm eq}} \hat{\rho}_{B {\rm eq}} e^{i \hat{H} \tau} \, . 
\end{align}
%------------------------------------------------------------------------------
}

\YU{
A naive definition of the operator of the ``local heat quantity" of the subsystem $A$ would be, 
%------------------------------------------------------------------------------
\begin{align}
\hat{Q}_{A}=\hat{S}_{A}/\beta_{A} = - \ln \hat{\rho}_{{A} {\rm eq}}/\beta_{A} \, . \label{eqn:heat_op}
\end{align}
%------------------------------------------------------------------------------
Precisely, Eq.~(\ref{eqn:heat_op}) is the operator of energy measured from the chemical potential minus the equilibrium free energy of the subsystem $A$. 
In thermodynamics, the heat is not a state function and is defined associated to a certain process. 
In our case, the process corresponds to the exchange of heat and electrons between the subsystem $A$ and the exterior, the subsystem $B$. 
Indeed, the time derivative of the average of the operator (\ref{eqn:heat_op}) is compatible with the commonly used definition of the heat flux (see Ref.~\onlinecite{EspositoPRB2015} for definitions of the heat), 
%------------------------------------------------------------------------------
\begin{align}
\frac{d}{dt} \langle \hat{Q}_{A}(t) \rangle = \dot{E}_{A}(t) - \mu_{A} \dot{N}_{A}(t)
\, .
\end{align}
%------------------------------------------------------------------------------
Here the averages of energy and particle currents are 
$\dot{E}_{A}(t)=-i {\rm Tr} \left( \hat{\rho}(\tau) [ \hat{H}_{A} , \hat{H} ] \right)$ 
and 
$\dot{N}_{A}(t)=-i {\rm Tr} \left( \hat{\rho}(\tau) [ \hat{N}_{A} , \hat{H} ] \right)$, respectively. 
}
Once we accept Eq.~(\ref{eqn:heat_op}), the projection operator can be written as \YU{(Appendix \ref{projection_operator})}, 
%------------------------------------------------------------------------------
\begin{align}
\hat{\Pi}_{S_{A}} = \frac{\Delta}{2 \pi} \int_{-\pi/\Delta}^{\pi/\Delta} d \chi e^{- i \chi S_{A}} \hat{\rho}_{{A} {\rm eq}}^{-i \chi} \, . 
\label{proj_sa}
\end{align}
%------------------------------------------------------------------------------
For simplicity, we assume that the \YU{dimensionless heat quantity} is discrete $S_{A}=\Delta n$, where $n$ is an integer. 
$\Delta$ is a small number, and we set $\Delta \to + 0$ at the end of the calculations. 
\YU{
Physically, this operation would correspond to taking the limit of large subsystem size in the end of calculations. 
As far as $\Delta >0$, $S_A \in (-\infty,\infty)$ and thus there would be no limitation on the bandwidth of the detector, i.e. the subsystem $A$. 
}

The reduced-density matrix after the projective measurement is $\hat{\rho}_A^\prime  = \sum_{S_A} P(S_A) \, \hat{\rho}_{A,S_A}$. 
We define the joint probability distribution function of self-information content and \YU{dimensionless heat quantity} as, 
%------------------------------------------------------------------------------
\begin{align}
P(I_A^\prime,S_A) = {\rm Tr}_A \left[ \hat{\Pi}_{S_A} \hat{\rho}_A \hat{\Pi}_{S_A} \delta ( I_A^\prime + \ln \hat{\rho}_A^\prime ) \right] \, . 
\label{jointpdf}
\end{align}
%------------------------------------------------------------------------------
%We also define the probability distribution function of conditional self-information as, 
\YU{
By using the joint probability distribution function, the probability distribution function of conditional self-information (\ref{condpdf}) can be written as, 
%------------------------------------------------------------------------------
%\begin{subequations}
\begin{align}
P_{S_A}(J) % &= {\rm Tr}_A \left[ \hat{\rho}_{A,S_A} \delta ( J + \ln \hat{\rho}_{A,S_A} ) \right] \label{condpdf} \\
&= P(I_A^\prime=J-\ln P(S_A),S_A)/P(S_A)
\, .
\label{cond_and_joint_pdf}
\end{align}
%\end{subequations}
%------------------------------------------------------------------------------
}
%For practical calculations, it is convenient to introduce the characteristic function or the information-generating function~\cite{Golomb1966,Guiasu1985}, the Fourier transform of the probability distribution function. 
The information-generating function of the joint probability distribution (\ref{jointpdf}) is, 
%------------------------------------------------------------------------------
\begin{align}
S_{1-i \xi}(S_A) =& \int d I_A^\prime e^{i \xi I_A^\prime} P(I_A^\prime,S_A) \nonumber \\ =& {\rm Tr}_A \left[ \left( \hat{\Pi}_{S_A} \hat{\rho}_A \hat{\Pi}_{S_A} \right)^{1-i \xi} \right] \, . 
\label{m_renyi}
\end{align}
%------------------------------------------------------------------------------
We call the parameter of Fourier transform $\xi$ the `counting field'. 
By performing the analytic continuation $1-i \xi \to M$, we obtain the R\'enyi entropy (\ref{renyi_q}). 
Because of their apparent similarity, we use the terms `information-generating function' and `R\'enyi entropy' interchangeably. 

We further perform the Fourier transform in terms of the \YU{dimensionless heat quantity}. 
By exploiting the expression (\ref{proj_sa}), we obtain, 
%------------------------------------------------------------------------------
\begin{align}
S_{1-i \xi}(\chi) = \sum_{S_A} e^{i \chi S_A} S_{1-i \xi}(S_A) = {\rm Tr}_A \left( \hat{\rho}_{A}^{\prime \; 1-i \xi} \hat{\rho}_{A {\rm eq}}^{-i \chi} \right) \, . \label{j_renyi}
\end{align}
%------------------------------------------------------------------------------
Once the R\'enyi entropy (\ref{j_renyi}) is obtained, the joint probability distribution is recovered by performing the inverse Fourier transform. 
In the present paper, we focus on the steady state realized in the limit of $\tau \to \infty$. 
\YU{In the presence of a finite affinity, the temperature difference or the chemical potential difference, the number of exchanged electrons grows linearly in the measurement time $\tau$. 
Since the information is conveyed by arrivals or non-arrivals of electrons, the self-information content as well as the heat quantity would grow in proportion to the measurement time $\tau$~\cite{YU2015,YU2017}. } 
Therefore, the inverse Fourier transform can be done within the saddlepoint approximation, 
%------------------------------------------------------------------------------
\begin{subequations}
\begin{align}
P(I_A^\prime,S_A) =& \frac{1}{2 \pi} \int d \xi e^{-i \xi I_A^\prime} S_{1-i \xi}(S_A) \label{eqn:ift1_exact}
\\
\approx & \exp \left[ \min_{i \xi \in {\mathbb R}} \left( \ln S_{1-i \xi}(S_A) - i \xi I_A^\prime \right) \right] \, , 
\label{eqn:ift1}
\end{align}
\end{subequations}
%------------------------------------------------------------------------------
which is the Legendre-Fenchel transform~\cite{Touchette2009}. 
For the joint probability distribution, we can perform the double Legendre transform; 
%------------------------------------------------------------------------------
\begin{align}
\ln P(I_A^\prime,S_A) \approx & \min_{i \xi, i \chi \in {\mathbb R}} \left( \ln S_{1-i \xi}(\chi) - i \xi I_A^\prime - i \chi S_A \right)
\, . 
\label{eqn:doubleLFT}
\end{align}
%------------------------------------------------------------------------------

Hereafter, we use $S_A$ and $Q_A$ interchangeably. 
The two quantities and corresponding counting fields, $\chi$ and $X$, are related as $Q_A=S_A/\beta_A$ and $X=\beta_A \chi$, respectively. 
The joint cumulant between self-information and heat quantity is obtained by a derivative of the information-generating function; 
%------------------------------------------------------------------------------
\begin{align}
\langle \! \langle I_A^{\prime \, \ell} Q_A^m \rangle \! \rangle = \left. \partial_{i \xi}^\ell \partial_{i X}^m  \ln S_{1- i \xi} (X) \right|_{\xi=X=0}
\, .
\label{iscum}
\end{align}
%------------------------------------------------------------------------------

\subsection{Universal relation and optimum capacity}
\label{sec:unirel}

The information-generating function of the probability distribution of conditional self-information (\ref{cond_and_joint_pdf}) is, 
%------------------------------------------------------------------------------
\begin{align}
S_{1-i \xi,S_A} = \int d J e^{i \xi J} P_{S_A}(J) = \frac{S_{1-i \xi}(S_A)} {S_{1}(S_A)^{1-i \xi}} \, . 
\label{c_renyi}
\end{align}
%------------------------------------------------------------------------------
The first derivative gives the von Neumann entropy~\cite{NC2000} 
$S( \hat{\rho} )=-{\rm Tr} \hat{\rho} \ln \hat{\rho}$
as,
%------------------------------------------------------------------------------
\begin{align}
\langle \! \langle J \rangle \! \rangle 
=
\left. \partial_{i \xi} \ln S_{1-i \xi,S_A} \right|_{i \xi=0}
=
S(\hat{\rho}_{A,S_A})
\, . 
\end{align}
%------------------------------------------------------------------------------
The information-generating function satisfies a Jarzynski equality~\cite{Jarzynski1997,Campisi2011} like universal relation, 
%------------------------------------------------------------------------------
\begin{align}
\left \langle e^J \right \rangle_{S_A} = & \int d J e^J P_{S_A}(J) 
=\int d J \, {\rm Tr}_A \left[ \delta (J-\hat{J}) \right]
\nonumber \\
=& S_{0,S_A} = {\rm rank} \, \hat{\rho}_{A,S_A} \nonumber \\ 
=& S_{0}(S_A) = {\rm rank} \left( \hat{\Pi}_{S_A} \hat{\rho}_A \hat{\Pi}_{S_A} \right) \, .  \label{jareq1}
\end{align}
%------------------------------------------------------------------------------
\YU{The last expression of the first line means the number of eigenvalues of the `entanglement Hamiltonian'~\cite{LiPRL2008,FrancisSong2012,Petrescu2014}, $\hat J = - \ln \hat{\rho}_{A,S_A}$. 
The last equation of the second line means the number of {\it positive} eigenvalues of the reduced density matrix $\hat{\rho}_{A,S_A}$. 
Therefore, $\left \langle e^J \right \rangle_{S_A}$ would represent the number of all possible many-body electron states in the subsystem $A$ for a given local dimensionless heat quantity $S_A$ occurring with positive probabilities. 
In general, the zeroth-order R\'enyi entropy gives the measure of the support set of a given probability density function, while the Shannon entropy gives the size of the effective support set~\cite{Cover2006}. 
}

\YU{
To proceed, we perform the Fourier transform of Eq.~(\ref{jareq1}), 
%------------------------------------------------------------------------------
$ S_0(X) = \sum_{Q_A} e^{i X Q_A} \left \langle e^J \right \rangle_{Q_A} = {\rm Tr}_A \left( \hat{\rho}_{A}^{\prime \; 0} \hat{\rho}_{A {\rm eq}}^{-i X/\beta_A} \right)$, 
%------------------------------------------------------------------------------
and then perform the inverse Fourier transform within the saddlepoint approximation, }
%------------------------------------------------------------------------------
\begin{align}
\ln \left \langle e^J \right \rangle_{Q_A} =& \ln \frac{\Delta}{2 \pi \beta_A} \int_{- \pi \beta_A/\Delta}^{\pi \beta_A/\Delta} d X e^{-i X Q_A} S_0(X) \nonumber \\ \approx& \min_{i \lambda \in  {\mathbb R} } \left( \ln S_0 \! \left( X= \lambda/\Delta E \right) - i \lambda n \right) \, , \label{ilft}
\end{align}
%------------------------------------------------------------------------------
where we introduced the heat quantity divided by the energy resolution $\Delta E=h/(2 N_{\rm ch} \tau)$, 
%------------------------------------------------------------------------------
$n = Q_A/\Delta E = \tau^2 N_{\rm ch} P_A/\pi$. 
%------------------------------------------------------------------------------
In Sec.~\ref{sec:optc}, we calculate the second line of Eq.~(\ref{ilft}) in the limit of $\beta_A \to \infty$ explicitly for a resonant-level model and reproduce the optimum capacity in the previous works~\cite{CavesRMP1994,BlencowePRA2000} Eq.~(\ref{eqn:pf}) as claimed by Eq.~(\ref{jareq}).

If $P_{Q_A}(J) \geq 0$, by exploiting Jensen's inequality and the universal relations, Eqs.~(\ref{jareq}) and (\ref{jareq1}), one can check that the average conditional self-information is bounded from above; 
%------------------------------------------------------------------------------
\begin{align}
\langle J \rangle_{Q_A} = S(\hat{\rho}_{A,Q_A}) \leq \ln {\rm rank} \, \hat{\rho}_{A,Q_A}  =  \tau \, C_{\rm opt}(P) \label{eqn:2ndlaw}
\, . 
\end{align}
%------------------------------------------------------------------------------

\YU{
Here we comment on the definition of the delta function in Eqs.~(\ref{condpdf}) and (\ref{jointpdf}) and the normalization condition of the joint probability distribution Eq.~(\ref{jointpdf}). 
By using the spectral decomposition of the reduced density matrix, 
%------------------------------------------------------------------------------
\begin{align}
\hat{\rho}_A^\prime = \sum_j \lambda_j \, | j \rangle \langle j | \, , 
\end{align}
%------------------------------------------------------------------------------
where $\lambda_j$ are non-negative eigenvalues, 
the delta function in Eqs.~(\ref{condpdf}) and (\ref{jointpdf}) is defined as, 
%------------------------------------------------------------------------------
\begin{align}
\delta \left(I_A' + \ln \hat{\rho}_A^\prime \right) = \sum_{j \in {\mathcal S} } |j \rangle \langle j| \, \delta \! \left(I_A' + \ln \lambda_j \right) \, . 
\end{align}
%------------------------------------------------------------------------------
Here, the summation is performed over the index $j$ associated with positive eigenvalues, ${\mathcal S} = \{ j : \lambda_j > 0 \}$. 
In the present paper, we will assume that the initial state of the subsystem $A$ is in a pure state. 
%------------------------------------------------------------------------------
\begin{align}
\rho_{A {\rm eq}} = | {\rm FS} \rangle \langle {\rm FS} |
\, , 
\end{align}
%------------------------------------------------------------------------------
where $| {\rm FS} \rangle$ is a unique ground-state many-body wave function (at the ground state, electrons fill up to the Fermi energy). 
Then 
%------------------------------------------------------------------------------
\begin{align}
\int dI_A' \int dS_A P(I'_A, S_A) = S_1(\chi=0)=\sum_{j \in {\mathcal S}} | \langle j | {\rm FS} \rangle |^2
\, , 
\label{eqn:normal_cond}
\end{align}
%------------------------------------------------------------------------------
which would not necessarily be 1. 
In the present paper, we will consider a specific model, a resonant-level model, and check the normalization condition through explicit calculations, see Eq.~(\ref{eqn:check_normal_cond}). 
}

\section{Multi-contour Keldysh technique}
\label{sec:mcKeldysh}

\subsection{Bulk contribution}

Let us calculate the R\'enyi entropy (\ref{j_renyi}) at the initial state $\tau=0$, in which the two subsystems are decoupled; 
%------------------------------------------------------------------------------
\begin{align}
s_{M}(\chi) = {\rm Tr}_B \left( \hat{\rho}_{A {\rm eq}}^{M-i \chi} \right) \, ,  
\label{eqn:s0}
\end{align}
%------------------------------------------------------------------------------
where we used $\hat{\rho}_A^\prime = \hat{\rho}_{A {\rm eq}}$. 
The operators of the Hamiltonian and particle number of the subsystem $A$ are, 
%------------------------------------------------------------------------------
\begin{subequations}
\begin{align}
\hat{H}_A &= 
\sum_{k} \epsilon_{A \, k} \hat{a}_{A k}^\dagger \hat{a}_{A k} 
\, ,  \\
\hat{N}_A &= 
\sum_{k} \hat{a}_{A k}^\dagger \hat{a}_{A k} 
\, . 
\end{align}
\end{subequations}
%------------------------------------------------------------------------------
Then the unperturbed part Eq.~(\ref{eqn:s0}) reads as, 
%------------------------------------------------------------------------------
\begin{align}
\ln s_M(\chi) =& \ln \frac{ {\rm Tr}_A e^{- (M - i \chi) \beta_A ( \hat{H}_A - \mu_A \hat{N}_A) } }{Z_{A {\rm eq} }^{M - i \chi}} 
\\ =&
\ln \frac{ \prod_{k} \left( 1+e^{-(M - i \chi ) \beta_A (\epsilon_{A k} - \mu_A)} \right) }{ \prod_{k} \left(1+e^{-\beta_A (\epsilon_{A k} - \mu_A)} \right)^{M - i \chi} }
\\ =&
\int d \omega {\mathcal N}_A(\omega) \ln \left( f_{A}^+(\omega)^{M-i \chi} + f_{A}^-(\omega)^{M-i \chi} \right) \, , 
\end{align}
%------------------------------------------------------------------------------
where ${\mathcal N}_A(\omega)=\sum_k \delta(\omega-\epsilon_{Ak})$ is the density of states (DOS) of the subsystem $A$. 
The electron (hole) distribution function is, 
%------------------------------------------------------------------------------
\begin{align}
f_A^\pm(\omega)
=
\frac{1} {1+e^{\pm \beta_A (\omega-\mu_A)}}
\, . 
\end{align}
%------------------------------------------------------------------------------

For further calculations, we assume the DOS is energy-independent 
${\mathcal N}_A(\omega)=V_A \rho_A$, 
where $V_A$ is the volume of subsystem $A$. 
The R\'enyi entropy is analytic around $\chi=0$ and $M=1$ and is proportional to the volume and specific heat of free electron gas~\cite{AshcroftMermin} $C_A=\rho_A \pi^2/(3 \beta_A)$ as~\cite{Nazarov2011}, 
%------------------------------------------------------------------------------
\begin{align}
\ln s_M(\chi)
\approx
\frac{V_A C_A}{2} \left( \frac{1}{M-i \chi}-M+i \chi \right)
\, . 
\label{eqn:bulk_renyi}
\end{align}
%------------------------------------------------------------------------------
In the limit of zero temperature $\beta_A \to \infty$, Eq.~(\ref{eqn:bulk_renyi}) becomes zero except at $M = i \chi$.

\subsection{Keldysh-generating function}

We adopt the replica trick to calculate the information-generating function Eq.~(\ref{j_renyi}). 
First, we calculate 
%------------------------------------------------------------------------------
\begin{subequations}
\begin{align}
S_{M}(\chi) = {\rm Tr}_A \left( \hat{\rho}_{A}^{\prime \; M} \hat{\rho}_{A {\rm eq}}^{-i \chi} \right) \, , 
\label{mrenyi}
\end{align}
%------------------------------------------------------------------------------
for a positive integer $M$ and then perform the analytic continuation back to $M \to 1-i \xi$. 
By utilizing expression of the projection operator (\ref{proj_sa}), the R\'enyi entropy becomes, 
%------------------------------------------------------------------------------
\begin{align}
S_M(\chi) =& \left( \frac{\Delta}{2 \pi} \right)^{M-1} \int^{\pi/\Delta}_{-\pi/\Delta} d \chi_M \cdots d \chi_1 \delta (\chi - \bar{\chi})
\nonumber \\ 
& \times S_M(\{ \chi_j \})
\, , 
\label{renyi_constraint}
\\
S_M(\{ \chi_j \}) =& {\rm Tr}_A \left[ { \hat{\rho}_{A {\rm eq}} }^{-i \chi_M} \hat{\rho}_A(\tau) \cdots { \hat{\rho}_{A {\rm eq}} }^{-i \chi_1} \hat{\rho}_A(\tau) \right]
\, , 
\label{eqn:renyi_multi}
\end{align}
\end{subequations}
%------------------------------------------------------------------------------
where $ \bar{\chi}=\sum_{j=1}^M \chi_j $. 
The operator of the ``local heat quantity" Eq.~(\ref{eqn:heat_op}) includes only the creation and annihilation operators acting locally on subsystem $A$. 
In this case, $\hat{\rho}_A(\tau)$ and $\hat{Q}_A$ are, in general, not commutative \YU{(Appendix~\ref{derivation_non_commutativity})}; 
%------------------------------------------------------------------------------
\begin{align}
[ \hat{\rho}_A(\tau), \hat{Q}_A ] =& {\rm Tr}_B \biggl( e^{-i \hat{H} \tau} \biggl[ \hat{\rho}_{A {\rm eq}} \hat{\rho}_{B {\rm eq}}, \hat{V} - e^{i \hat{H} \tau} \hat{V} e^{-i \hat{H} \tau} \biggl] 
\nonumber \\
& \times 
e^{i \hat{H} \tau} \biggl)
\neq 0 \, . 
\label{eqn:noncommutativity}
\end{align}
%------------------------------------------------------------------------------
Therefore, we must deal with multiple integrals over $\chi_j$ in Eq.~(\ref{renyi_constraint}). 
This situation contrasts with the {\it local} particle number constraint~\cite{YU2017}, in which (under certain conditions) the coherence between sectors of different particle numbers vanishes $ [\hat{\rho}_A, \hat{N}_A] =0$ (see Eq.~(26) and Appendix A of Ref.~\onlinecite{YU2017}) and thus the $M$-multiple integral is reduced to a single integral. 

\begin{figure}[hb]
\includegraphics[width=0.9 \columnwidth]{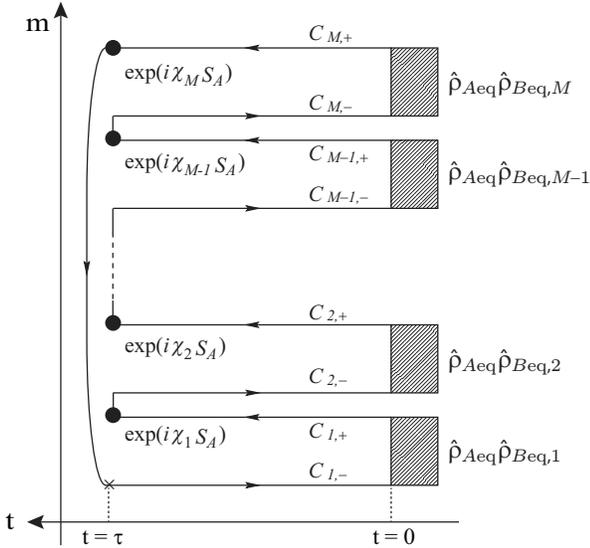}
\caption{
Multi-contour $C$ consisting of $M$ normal Keldysh contours, 
$C_1, \cdots, C_M$. 
A cross at $t=\tau$ on the lower branch of the first Keldysh contour $C_{1 , -}$ represents a starting point. 
The contour goes to $\hat{\rho}_{A {\rm eq}} \hat{\rho}_{B {\rm eq}, 1}$ at $t=0$ along $C_{1 , -}$ and returns to $t=\tau$ along $C_{1 , +}$. 
Then it connects to $t=\tau$ on the lower branch of the second Keldysh contour $C_{2 , -}$. 
The contour goes repeatedly until it reaches $t=\tau$ on $C_{M , +}$. 
Then it goes back to the starting point $t=\tau$ on $C_{1 , -}$. 
Shaded boxes are $M$ replicas of the initial equilibrium density matrix $\hat{\rho}_{A {\rm eq}} \hat{\rho}_{B {\rm eq}, m}$ ($m=1,\cdots,M$). 
Solid circles on $t=\tau_{m+}$ represent operators $\exp(i \chi_m \hat{S}_A)$. 
}
\label{mkc}
\end{figure}

Equation~(\ref{eqn:renyi_multi}) is expressed as the Keldysh partition function defined on a multi-contour~\cite{Nazarov2011,Ansari2015,YU2015,YU2017,Ansari2015b,YU2018fqmt}. 
The multi-contour $C$ is a sequence of $M$ normal Keldysh contours, $C_1, \cdots, C_M$ (Fig.~\ref{mkc}). 
We introduce $M$ replicas of creation and annihilation operators of the subsystem $B$, $\hat{a}_{B k} (\hat{a}_{B k}^\dagger) \to \hat{a}_{B k,m} (\hat{a}_{B k,m}^\dagger)$ living on the $m$th Keldysh contour $C_m$ ($m=1,\cdots,M$). 
The operators of the Hamiltonian and the number of particles of subsystem $B$ are replicated as $\hat{H}_{B} \to \hat{H}_{B,m}$ and $\hat{N}_{B} \to \hat{N}_{B,m}$, respectively. 
In addition, the operator of the coupling is replicated as $\hat{V} \to \hat{V}_m$. 
Then the R\'enyi entropy (\ref{eqn:renyi_multi}) is written in the form of the Keldysh partition function; 
%------------------------------------------------------------------------------
\begin{subequations}
\begin{align}
S_M(\{ \chi_j \}) &= \left \langle \hat{T}_C e^{-i \int_C dt \hat{V}(t)_I + i \sum_{m=1}^M \chi_m \hat{S}_A(\tau_{m+})_I } \right \rangle_M \nonumber \\ & \times s_M(\bar{\chi}) \, , 
\label{eqn:keldysh_partition_function}
\end{align}
%------------------------------------------------------------------------------
where $\hat{T}_C$ is the contour-ordering operator~\cite{YU2015,YU2017}. 
The operators in the interaction picture at time $t_{m \pm}$ on the upper (lower) branch of $m$th Keldysh contour are, e.g., 
$V(t_{m \pm})_I=e^{i (\hat{H}_{A}+\hat{H}_{B,m}) t} \hat{V}_m e^{-i (\hat{H}_{A}+\hat{H}_{B,m}) t}$. 
The average is, 
%------------------------------------------------------------------------------
\begin{align}
\langle \hat{ {\mathcal O} } \rangle_M
=
{\rm Tr} \left( \hat{ {\mathcal O} } \hat{\rho}_{A {\rm eq}} \hat{\rho}_{B {\rm eq} , M} \cdots \hat{\rho}_{A {\rm eq}} \hat{\rho}_{B {\rm eq} , 1} \right)
/s_M(\bar{\chi})
\, , 
\end{align}
\end{subequations}
%------------------------------------------------------------------------------
where $s_M$ is the unperturbed part of the R\'enyi entropy~(\ref{eqn:s0}). 
The density matrix of subsystem $B$ is also replicated as $\hat{\rho}_{B {\rm eq}, m}$. 
The trace is performed over the Hilbert space of subsystem $A$ and $M$ replicas of subsystem $B$. 
The result Eq.~(\ref{eqn:keldysh_partition_function}) is Eq.~(46) in Ref.~\onlinecite{YU2017} replaced $\hat{N}_A$ with $\hat{S}_A$. 
For detailed derivations, see Ref.~\onlinecite{YU2017}.

%------------------------------------------------------------------------------
%----------------------------------------------------------

\YU{

\subsection{Multi-contour Keldysh Green functions}

Here we illustrate the multi-contour Keldysh Green function for a simple model, 
%------------------------------------------------------------------------------
\begin{align} \hat{H}_{r}=\epsilon_{r} \hat{a}_{r}^\dagger \hat{a}_{r} \, , \;\;\;\; (r=A,B) \, .  \end{align}
%------------------------------------------------------------------------------
We relegate details to Appendix \ref{sec:mckgf} and summarize definitions. 
A multi-contour Keldysh Green function of subsystem $A$ is a contour-ordered correlation function of $\hat{a}_A^\dagger$ on $C_{m' , s'}$ and $\hat{a}_A$ on $C_{m , s}$; 
%------------------------------------------------------------------------------
\begin{subequations}
\begin{align}
g_{A}^{ \{ \chi_j \} }(t_{ms},t'_{m's'}) =& g_{A}^{ \{ \chi_j \}, ms, m's' }(t,t') \nonumber \\
=& -i \left \langle \hat{T}_C \hat{a}_{A}(t_{ms})_I \hat{a}_{A}^\dagger(t_{m's'}')_I 
\right. \nonumber \\ &\times \left. 
e^{i \sum_{j'=1}^M \chi_{j'} \hat{S}_A(\tau_{j'+})_I} \right \rangle_M
\, . 
\label{gf}
\end{align}
%------------------------------------------------------------------------------
This is a ($ms,m's'$) component of a $2M \times 2M$ Keldysh Green function matrix ${\bm g}_{A}$ [see Eq.~(\ref{mmckgf_t}) for explicit expressions of components]. 
It is convenient to introduce the Fourier transform in time; 
%------------------------------------------------------------------------------
\begin{align}
{\bm g}_{A}^{ \{ \chi_j \} }(\omega) &= \int d (t-t') e^{i \omega (t-t')} {\bm g}_{A}^{ \{ \chi_j \} }(t,t') \nonumber \\
&= {\bm U}(\{ \delta \chi_j \},\omega)^\dagger {\bm g}_{A}^{ \bar{\chi} }(\omega) {\bm U}(\{ \delta \chi_j \},\omega)
\, , 
\label{eqn:unimgf}
\end{align}
%------------------------------------------------------------------------------
which is separated into a matrix ${\bm g}_{A}^{ \bar{\chi} }$ depending only on the average of the counting fields $\bar{\chi}=\sum_{m=1}^M \chi_m$ and a diagonal unitary matrix ${\bm U}$ depending only on fluctuations $\delta \chi_j = \chi_j - \bar{\chi}/M$ ($j=1,\cdots,M-1$). 
A ($ms,m's'$) component of the diagonal unitary matrix is, 
%------------------------------------------------------------------------------
\begin{align}
\left[ {\bm U} \right]_{ms,m's'} = e^{-i \phi_m(\omega)} \delta_{m,m'} \delta_{s,s'} \, . 
\label{eqn:dia_uni_mat}
\end{align}
\end{subequations}
%------------------------------------------------------------------------------
The phase $\phi_m$ is the accumulation of the fluctuations; 
$\phi_m(\omega)=\sum_{j=1}^{m-1} \delta \chi_{j} s_A(\omega)$ for $m=2,\cdots M$ and $\phi_1(\omega)=0$. 
We introduced the dimensionless heat quantity associated with a single electron excitation, $s_A(\omega)= \beta_A (\omega - \mu_A)$. 
The matrix ${\bm g}_{A}^{ \bar{\chi} }(\omega)$ is a block skew-circulant matrix, see Eqs.~(\ref{mmckgf})-(\ref{eqn:g_a_off_diag}).

Similarly, a multi-contour Keldysh Green function of subsystem $B$ is introduced. 
It is nonzero only when $\hat{a}_{B,m}$ and $\hat{a}_{B,m'}^\dagger$ are on the same normal Keldysh contour $m=m'$; 
%------------------------------------------------------------------------------
\begin{align}
g_{B}(t_{ms},t'_{m's'}) =& -i {\rm Tr}_{B,m} \left [ T_{C_m} \hat{a}_{B}(t_{ms})_I \hat{a}_{B}^\dagger(t_{m s'}')_I
\right. \nonumber \\ &\times \left. 
\hat{\rho}_{B {\rm eq},m} \right ] \delta_{m,m'} \, . 
\label{gfB}
\end{align}
%------------------------------------------------------------------------------
In the following calculations, we will use its Fourier transform in time, see Eqs.~(\ref{eqn:mmckgfB}) and (\ref{eqn:mmckgfB_omega}). 
}

\section{Resonant-level model}
\label{sec:srlm}

We consider the spinless resonant-level model [Fig.~\ref{setup} (b)]. 
We bipartition the system and regard the left lead as subsystem $A$ and the dot and right lead as subsystem $B$. 
The Hamiltonians of the two subsystems are, 
%------------------------------------------------------------------------------
\begin{subequations}
\begin{align}
\hat{H}_A =& \sum_{k} \epsilon_{Lk} \hat{a}_{L k}^\dagger \hat{a}_{L k} \, ,
\\
\hat{H}_B =& \sum_{k} \epsilon_{Rk} \hat{a}_{R k}^\dagger \hat{a}_{R k} + \epsilon_D \hat{d}^\dagger \hat{d} \, , 
\end{align}
%------------------------------------------------------------------------------
where $\hat{a}_{r k}$ annihilates an electron with wave number $k$ in the lead $r$ and $\hat{d}$ annihilates an electron in the quantum dot. 
Here $\epsilon_D$ is the energy of a localized level in the dot and $\epsilon_{rk}$ is the energy of the electron in the lead $r$. 
The coupling between the two subsystems is described by the tunnel Hamiltonian; 
%------------------------------------------------------------------------------
\begin{align}
\hat{V} = \sum_{r=L,R} \sum_{k} J_r \hat{d}^\dagger \hat{a}_{r k} + {\rm H.c.}
\end{align}
\end{subequations}
%------------------------------------------------------------------------------
The particle number operators in subsystems $A$ and $B$ are, 
%------------------------------------------------------------------------------
$\hat{N}_A = \hat{N}_L = \sum_{k} \hat{a}_{L k}^\dagger \hat{a}_{L k}$ 
and 
$\hat{N}_B = \hat{N}_R + \hat{N}_D = \sum_{k} \hat{a}_{R k}^\dagger \hat{a}_{R k} +\hat{d}^\dagger \hat{d}$, 
%------------------------------------------------------------------------------
respectively. 
\YU{
The inverse temperatures (chemical potentials) of the left and right leads are $\beta_L=\beta_A$ ($\mu_L=\mu_A$) and $\beta_R=\beta_B$ ($\mu_R=\mu_B$). 
As for the initial isolated dot, one may choose an arbitrary density matrix, since, in the steady state, the occupation of the dot level is governed by the electron distribution of the leads and is independent of the initial density matrix of the dot. 
Therefore, we assume that the initial density matrix of the dot possesses the same form as the equilibrium density matrix (\ref{eqn:Z_eq}) and is characterized by two auxiliary parameters, the `inverse temperature' $\beta_D$ and the `chemical potential' $\mu_D$. 
This form is convenient since it enables us to utilize the Bloch-De Dominicis theorem (Appendix A of Ref.~\onlinecite{YU2015}). 
As demonstrated in Appendix \ref{sec:dotgfm}, the parameters $\beta_D$ and $\mu_D$ disappear in the course of calculations and the final result, Eq.~(\ref{eqn:modrenyi0k}), is independent of the two parameters. 
}

The Keldysh partition function (\ref{eqn:keldysh_partition_function}) can be calculated by exploiting the linked cluster expansion~\cite{YU2015,YU2017}. 
In the limit of long measurement time, the leading contribution is proportional to $\tau$; 
%------------------------------------------------------------------------------
\begin{align}
\ln \frac{S_{M}( \{ \chi_j \} )}{s_{M}( \bar{\chi} )} \approx 
\tau \int \frac{d \omega}{2 \pi} \ln \frac{ {\rm det} \left[ {\bm G}_{D}^{ \{ \chi_j \}  } (\omega)^{-1} \right]}{ {\rm det} \left[ {\bm g}_{D}(\omega)^{-1} \right] }
\, ,
\label{eqn:linked_cluster_exp}
\end{align}
%------------------------------------------------------------------------------
where the full Green function matrix of the dot is, 
%------------------------------------------------------------------------------
\begin{align}
{ {\bm G}_{D}^{ \{ \chi_j \} } }^{-1} =& {\bm g}_{D}^{-1} - ({\bm 1} \otimes {\bm \tau}_3) \sum_k \left( J_L^2 {\bm U}^\dagger {\bm g}_{Lk}^{ \bar{\chi}} {\bm U} \right. \nonumber \\ & \left. + J_R^2 {\bm g}_{Rk} \right) ({\bm 1} \otimes {\bm \tau}_3) \, . 
\label{eqn:fgfm}
\end{align}
%------------------------------------------------------------------------------
Here ${\bm g}_{Lk}^{ \bar{\chi}}$ is obtained from ${\bm g}_{A}^{ \bar{\chi}}$ [Eq.~(\ref{mmckgf})] by replacing $\epsilon_A$ with $\epsilon_{Lk}$. 
Similarly, ${\bm g}_{D(Rk)}$ is obtained from ${\bm g}_{B}$ [Eq.~(\ref{eqn:mmckgfB})] by replacing $\epsilon_B$ with $\epsilon_{D(Rk)}$. 
\YU{
The diagonal unitary matrix ${\bm U}$ was introduced in Eq.~(\ref{eqn:dia_uni_mat}). 
Equation~(\ref{eqn:fgfm}) can be written as, ${ {\bm G}_{D}^{ \{ \chi_j \} } }^{-1} = {\bm U}^\dagger { {\bm G}_{D}^{ \bar{\chi} } }^{-1} {\bm U}$, where 
%------------------------------------------------------------------------------
\begin{align}
{ {\bm G}_{D}^{ \bar{\chi}} }^{-1} = {\bm g}_{D}^{-1} - ({\bm 1} \otimes {\bm \tau}_3) \sum_k \left( J_L^2 {\bm g}_{Lk}^{ \bar{\chi}} + J_R^2 {\bm g}_{Rk} \right) ({\bm 1} \otimes {\bm \tau}_3) \, . 
\end{align}
%------------------------------------------------------------------------------
By exploiting the property of the determinant, 
%------------------------------------------------------------------------------
\begin{align}
{\rm det} \left[ { {\bm G}_{D}^{ \{ \chi_j \} } }^{-1} \right] = {\rm det} \left[ {\bm U}^\dagger { {\bm G}_{D}^{ \bar{\chi} } }^{-1} {\bm U} \right] = {\rm det} \, { {\bm G}_{D}^{ \bar{\chi} } }^{-1} 
\, ,
\end{align}
%------------------------------------------------------------------------------
we observe that the phase $\phi_m(\omega)$ cancels and thus Eq.~(\ref{eqn:linked_cluster_exp}) depends only on the average $\bar{\chi}$. 
}
This cancellation originates from the energy conservation in the steady state~\cite{SU2008}. 
It implies that in the steady state, $\hat{\rho}_A$ and $\hat{Q}_A$ commute; see also Sec.~\ref{sec:couint}. 

\YU{Since ${\bm g}_{Lk}^{ \bar{\chi}}$ is a block skew-circulant matrix, it is block-diagonalized by the discrete Fourier transform (\ref{dft}). 
Then Eq.~(\ref{renyi_constraint}) is calculated as,}
%------------------------------------------------------------------------------
\begin{align}
\ln \frac{S_{M}(\chi)}{s_{M}(\chi)} \approx & \sum_{\ell=0}^{M-1} \tau \int \frac{d \omega}{2 \pi} \ln \frac{ {\rm det} \left[ {\mathcal G}_{D}^{ \lambda_\ell-\chi s_A(\omega)/M } (\omega)^{-1} \right]}{ {\rm det} \left[ {\mathbf g}_{D}(\omega)^{-1} \right] }
\nonumber \\
=& \tau \sum_{\ell =0}^{M-1} {\mathcal F}_G(\lambda_\ell-\chi s_A(\omega)/M) \, , 
\label{maires}
\end{align}
%------------------------------------------------------------------------------
where the full Green function matrix in the $2 \times 2$ normal Keldysh space is, 
%------------------------------------------------------------------------------
\begin{align}
{ {\mathcal G}_{D}^{ \lambda } }^{-1} = {\mathbf g}_{D}^{-1} - {\bm \tau}_3 \sum_k \left( J_L^2 {\mathbf g}_{Lk}^{\lambda} + J_R^2 {\mathbf g}_{Rk} \right) {\bm \tau}_3 \, . 
\label{eqn:dyson_22_keldysh}
\end{align}
%------------------------------------------------------------------------------
The free Green functions ${\mathbf g}$ are $2 \times 2$ matrices [see Eqs.~(\ref{eqn:free_Keldysh_gf}) and (\ref{eqn:mmckgfB_omega})]. 
The solution to this Dyson equation is given by Eq.~(\ref{hgf}) in Appendix \ref{sec:dotgfm}. 
The function ${\mathcal F}_{G}$ is related to the scaled cumulant-generating function of the full counting statistics, 
%------------------------------------------------------------------------------
\begin{align}
{\mathcal F}_{G}(\lambda) &= \frac{N_{\rm ch}}{2 \pi} \int d \omega \ln \Omega_{1,\lambda}( \omega ) \, , \;\;\;\; N_{\rm ch}=1 \, , 
\nonumber \\
\Omega_{M, \lambda}( \omega ) &= \frac{ \tilde{f}_L^+(\omega)^M + \tilde{f}_L^-(\omega)^M e^{i \lambda} }{ {f}_L^+(\omega)^M + {f}_L^-(\omega)^M e^{i \lambda} }\, , \label{scgffree}
\end{align}
%------------------------------------------------------------------------------
where we subtracted a trivial constant to satisfy the normalization condition ${\mathcal F}_{G}(0)=0$. 
We introduced the effective electron (hole) distribution function 
%------------------------------------------------------------------------------
$ \tilde{f}_L^\pm(\omega) = {\mathcal T}(\omega) f_R^\pm(\omega) + {\mathcal R}(\omega) f_L^\pm(\omega) $, 
%------------------------------------------------------------------------------
where ${\mathcal T}(\omega)$ is the transmission probability and ${\mathcal R}(\omega) = 1-{\mathcal T}(\omega)$ is the reflection probability; 
%------------------------------------------------------------------------------
\begin{align}
{\mathcal T}(\omega) = \frac{\Gamma_L \Gamma_R}{(\omega - \epsilon_D)^2 + \Gamma^2/4} \, , \;\;\;\; \Gamma = \Gamma_L + \Gamma_R \, . 
\end{align}
%------------------------------------------------------------------------------
The coupling strength between the quantum dot and the lead $r$, 
$\Gamma_r = 2 \pi \sum_k J_r^2 \delta (\omega - \epsilon_{rk})$, 
is assumed to be energy independent. 
After we perform the summation over $\ell$ in Eq.~(\ref{maires}) (Appendix \ref{sec:summation}), we obtain, 
%------------------------------------------------------------------------------
\begin{align}
\ln \frac{ S_{M}(\chi) }{s_{M}(\chi)} \approx \frac{\tau N_{\rm ch}}{2 \pi} \int d \omega \, \ln \Omega_{M, -\chi s_A(\omega)}(\omega) \, . \label{eqn:renyi11}
\end{align}
%------------------------------------------------------------------------------
The above results are modifications of those obtained in Refs.~\onlinecite{YU2015} and \onlinecite{YU2017}. 
Equation (\ref{maires}) is Eq.~(64) in Ref.~\onlinecite{YU2017}; $\chi$ is replaced by $\chi s_A(\omega)$. 
Expressions in Eq.~(\ref{scgffree}) are Eqs.~(53) and (54) in Ref.~\onlinecite{YU2015}. 
Technical details can be found in these works.

The order in which the zero temperature limit and the analytic continuation are taken is important when we consider universal relations associated with the R\'enyi entropy of order zero~\cite{YU2015,YU2017}. 
Here we take the zero temperature limit only for subsystem $A$, 
$\beta_A \to \infty$, while keeping $M$ as a positive integer and the counting field $\beta_A \chi=X$ finite. 
By setting $f_L^+(\omega)=\theta(\mu_A-\omega)$, which is the Heaviside step function, Eq.~(\ref{eqn:renyi11}) becomes, 
%------------------------------------------------------------------------------
\begin{align}
\ln S_M(X) =& \frac{\tau N_{\rm ch}}{2 \pi} \int_{\mu_A}^\infty d \omega \ln \left[ \left( {\mathcal T}(\omega) f_R^+(\omega) \right)^M e^{i X (\omega-\mu_A)} 
\right. \nonumber \\ & \left. 
+ \left(1- {\mathcal T}(\omega) f_R^+(\omega) \right)^M \right]
\nonumber \\
& + \frac{\tau N_{\rm ch}}{2 \pi} \int_{-\infty}^{\mu_A} d \omega \ln \left[ \left( 1 - {\mathcal T}(\omega) f_R^-(\omega) \right)^M 
\right. \nonumber \\ & \left. 
+ \left( {\mathcal T}(\omega) f_R^-(\omega) \right)^M e^{-i X (\omega-\mu_A)} \right] \, . 
\label{eqn:modrenyi0k}
\end{align}
%------------------------------------------------------------------------------
The two terms on the RHS of the equation correspond to the electron and hole contributions. 
By performing the analytical continuation $M \to 1- i \xi$, we obtain the information-generating function. 
\YU{
We check that Eq.~(\ref{eqn:modrenyi0k}) satisfies, 
%------------------------------------------------------------------------------
\begin{align}
\ln S_{1}(X=0) = 0 \, , 
\label{eqn:check_normal_cond}
\end{align}
%------------------------------------------------------------------------------
and thus the joint probability distribution function is properly normalized to 1, see Eq.~(\ref{eqn:normal_cond}). 
}

\section{Optimum capacity}
\label{sec:opt_cap}

\subsection{Averages}

From Eq.~(\ref{eqn:modrenyi0k}), by exploiting Eq.~(\ref{iscum}), the average of the self-information is evaluated;
%------------------------------------------------------------------------------
\begin{align}
\langle \! \langle I_A' \rangle \! \rangle =& \frac{\tau N_{\rm ch}}{2 \pi} \int_{\mu_A}^\infty d \omega H_2 \left( {\mathcal T}(\omega) f_R^+(\omega) \right) \nonumber \\ & + \frac{\tau N_{\rm ch}}{2 \pi} \int_{-\infty}^{\mu_A} d \omega H_2 \left( {\mathcal T}(\omega) f_R^-(\omega) \right) \, , 
\label{eqn:av_I_A}
\end{align}
%------------------------------------------------------------------------------
where we introduced the binary entropy $H_2(x)=-x \ln x - (1-x) \ln (1-x)$. 
The first and second terms on the RHS correspond to electron and hole contributions, respectively. 
The integrand $H_2 \left( {\mathcal T} f_R^+ \right)$ is the entropy of the receiver side. 
It corresponds to $H(B)$ of Eq. (21) in Ref.~\onlinecite{Akkermans2009} and Eq.~(12) of Ref.~\onlinecite{Sivan1986}. 
The average heat quantity in the left reservoir is 
%------------------------------------------------------------------------------
\begin{align}
\langle \! \langle Q_A \rangle \! \rangle =& \frac{\tau N_{\rm ch}}{2 \pi} \int d \omega (\omega - \mu_A) {\mathcal T}(\omega) ( f_R^+(\omega) - {f}_L^+(\omega) ) \, , 
\label{eqn:av_Q_A}
\end{align}
%------------------------------------------------------------------------------
which corresponds to the average signal power, Eq.~(15) in Ref.~\onlinecite{Akkermans2009}. 
%See also Eq.~(13) in Ref.~\onlinecite{Sivan1986}. 

\subsection{Optimum capacity and integer partitions}
\label{sec:optc}

Let us take $M \to 0$ of Eq.~(\ref{eqn:modrenyi0k}) while keeping the inverse temperature $\beta_B$ finite. 
For ${\mathcal T}(\omega) > 0$, $({\mathcal T} f_R^\pm)^0 =({\mathcal R}+{\mathcal T} f_R^\pm)^0 =1$. 
For ${\mathcal T}(\omega) = 0$, $({\mathcal T} f_R^\pm)^0=0$, and $({\mathcal R}+{\mathcal T} f_R^\pm)^0 =1$. 
Therefore, in this limit, Eq.~(\ref{eqn:modrenyi0k}) is independent of the details of the setup and depends only on the statistics of particles. 

We consider the band-limited channel; we introduce a finite bandwidth, i.e., a high-frequency cutoff $\omega_{\max}$ and a low-frequency cutoff, or a gap, $\omega_{\min}>0$. 
As is observed from Eq.~(\ref{eqn:modrenyi0k}), electrons ($\omega>\mu_A$) and holes ($\omega<\mu_A$) contribute in the same way. 
Therefore, in the following, when only electrons contribute, i.e., $\omega_{\min}< \omega < \omega_{\max}$, we set $N_{\rm ch}=1/2$. 
When electrons and holes contribute, i.e., $\omega_{\min}< |\omega| < \omega_{\max}$, we set $N_{\rm ch}=1$. 
The R\'enyi entropy of order zero is, 
%------------------------------------------------------------------------------
\begin{align}
\ln S_0(X=\lambda/\Delta E) &= \frac{1}{\Delta E} \int_{\omega_{\min}}^{\omega_{\max}} d \omega \ln \left( e^{i X \omega}+ 1 \right) \nonumber \\ &= \int^{j_{\max}}_{j_{\min}} d j \ln \left( 1+e^{i \lambda j} \right) \, , \label{gefuip}
\end{align}
%------------------------------------------------------------------------------
where $j= \omega/\Delta E$. 
The energy resolution is $\Delta E=h/(2 N_{\rm ch} \tau)$, see Eq.~(\ref{ilft}). 
\YU{It is an approximation of the logarithm of the generating function for partitions~\cite{Andrews2004}; $\prod_{j \in {\mathcal S}}(1+e^{i \lambda j})=\sum_{n \geq 0} p(n| {\rm distinct \, parts \, in } \, {\mathcal S}) \, e^{i n \lambda}$, where ${\mathcal S}=\{ j_{\min}, j_{\min}+1, \cdots, j_{\max} \}$. 
The {\it partition function}~\cite{Andrews2004} $p(n| {\rm distinct \, parts \, in } \, {\mathcal S})$ stands for the number of {\it integer partitions} of a given integer $n$ into distinct elements of the set ${\mathcal S}$. The {\it integer partition} of $n$ is a way of writing $n$ as the sum of positive integers. }
By exploiting Eq.~(\ref{ilft}), we obtain, 
%------------------------------------------------------------------------------
\begin{align}
\left \langle e^J \right \rangle_{Q_A} \approx p( Q_A/\Delta E| {\rm distinct \, parts \, in } \, {\mathcal S}) \, , 
\label{eqn:pf}
\end{align}
%------------------------------------------------------------------------------
which is $\exp \left( \tau C_{\rm opt}(P_A) \right)$ according to the previous quantum information theory approach in Ref.~\onlinecite{BlencowePRA2000} (see Eq.~(\ref{optc_caves_fermions}) in Sec.~\ref{sec:previous_theories}).

The result presented above verifies our main claim Eq.~(\ref{jareq}). 
However, precisely speaking, there are differences. 
The previous works~\cite{CavesRMP1994,BlencowePRA2000} treated dispersionless channels. 
Our result is derived from a microscopic Hamiltonian and can be extended to channels with arbitrary dispersion. 
We present more detailed comparisons in Sec.~\ref{sec:previous_theories}. 

The integral in Eq.~(\ref{gefuip}) can be done analytically; 
%------------------------------------------------------------------------------
\begin{align}
\ln S_0(X=\lambda/\Delta E) = \frac{{\rm Li}_2 \left( -e^{i \lambda j_{\min}} \right) - {\rm Li}_2 \left( -e^{i \lambda j_{\max}} \right)}{i \lambda} \, 
\end{align}
%------------------------------------------------------------------------------
where the dilogarithm function is, 
%------------------------------------------------------------------------------
\begin{align}
{\rm Li}_2(x)= \sum_{k=1}^\infty \frac{x^k}{k^2} = \int_x^0 dz \frac{\ln(1-z)}{z} \, .
\end{align}
%------------------------------------------------------------------------------
For the narrowband case, when the bandwidth $2 \pi B = \omega_{\max} - \omega_{\min}$ and the frequency $2 \pi f=(\omega_{\max} + \omega_{\min})/2$ satisfy $B \ll f$, the generating function is approximately, 
%------------------------------------------------------------------------------
%\begin{align}
$ \ln S_0(X=\lambda/\Delta E) \approx 2 \tau N_{\rm ch} B \ln \left( 1+e^{i 2 \tau N_{\rm ch} f \lambda } \right) $. 
%\end{align}
%------------------------------------------------------------------------------
Then by substituting it into Eqs.~(\ref{ilft}) and (\ref{jareq}), we obtain, 
%------------------------------------------------------------------------------
\begin{subequations}
\begin{align}
C_{\rm opt}(P_A) \approx 2 N_{\rm ch} B H_2(P_A/(2 N_{\rm ch} h f B)) \, . 
\label{eqn:cnb}
\end{align}
%------------------------------------------------------------------------------
In the particle-like regime, $P_A \ll 2 N_{\rm ch} h f B$, where the signal power is small and the particle nature of an electron is prominent, 
%------------------------------------------------------------------------------
\begin{align}
C_{\rm opt}(P_A) \approx \frac{P_A}{hf} \ln \frac{2 N_{\rm ch}  hfB}{P_A} \, . 
\end{align}
\end{subequations}
%------------------------------------------------------------------------------
Here $P_A/(hf)$ is the rate of transmission of signal quanta. 
The argument of the logarithm $2 N_{\rm ch} h f B/P_A$ means the maximum number of distinguishable modes per signal quantum. 
For $N_{\rm ch}=1/2$, the expression is formally compatible with Eq.~(2.22) in Ref.~\onlinecite{CavesRMP1994} that was obtained for bosons.

For the fermionic band-limited channel, the power of the signal is bounded from above. 
Let us set $\omega_{\rm min} = 0$. 
The maximum of the heat quantity is, 
%------------------------------------------------------------------------------
\begin{subequations} 
\begin{align} 
\frac{ Q_{A \, {\rm max}} }{\Delta E} = \lim_{i \lambda \to \infty} \frac{\ln S_0(X=\lambda/\Delta E)}{i \lambda} = \frac{j_{\max}^2}{2} \, , 
\label{eqn:n_max}
\end{align}
%------------------------------------------------------------------------------
where we utilized the Legendre duality~\cite{Touchette2009} and the fact that a rare event associated with the maximum is realized in the limit of $i \lambda \to \infty$. 
The maximum power is, 
%------------------------------------------------------------------------------
\begin{align}
P_{A \, {\max}} = \frac{ Q_{A \, {\rm max}} }{\tau} = \frac{2 N_{\rm ch}}{h} \int_0^{\omega_{\max}} \omega d \omega \, 
\label{eqn:P_max}
\end{align}
\end{subequations} 
%------------------------------------------------------------------------------
which is the Landauer formula of heat current for perfect transmission.

\YU{Let us turn our attention to the wideband channel, $j_{\max} \to \infty$. 
As long as the inverse Fourier transform is performed within the saddlepoint approximation, see Eq.~(\ref{ilft}), it is sufficient to analyze the generating function (\ref{gefuip}) for pure imaginary $\lambda$. 
The integral in Eq.~(\ref{gefuip}) can be done for $i \lambda <0$ and we obtain, }
%------------------------------------------------------------------------------
\begin{subequations} 
\begin{align}
\ln S_0(X=\lambda/\Delta E) \approx - \frac{\pi^2}{12 i \lambda} 
%\, , \;\;\;\; (i \lambda <0) 
\, .
\end{align}
%------------------------------------------------------------------------------
%For the wideband channel, $j_{\max} \to \infty$, the generating function (\ref{gefuip}) is approximately, 
%------------------------------------------------------------------------------
%\begin{subequations}  \begin{align} \ln S_0(X=\lambda/\Delta E) \approx - \frac{\pi^2}{12 i \lambda} \, , \;\;\;\; (i \lambda <0) \, . \end{align}
%------------------------------------------------------------------------------
Then by substituting it into Eq.~(\ref{ilft}) and by using Eq.~(\ref{jareq}), we reproduce the optimum capacity of the wideband channel, Eq.~(\ref{eqn:cwb}); 
%------------------------------------------------------------------------------
\begin{align}
C_{\rm opt} (P_A) \approx \frac{\pi}{\tau} \sqrt{\frac{Q_A/\Delta E}{3}} = C_{\rm WB}(P_A) \, , \label{eqn:cwb1}
\end{align}
\end{subequations} 
%------------------------------------------------------------------------------
where $P_A>0$. 

\begin{figure}[h]
\includegraphics[width=0.65 \columnwidth]{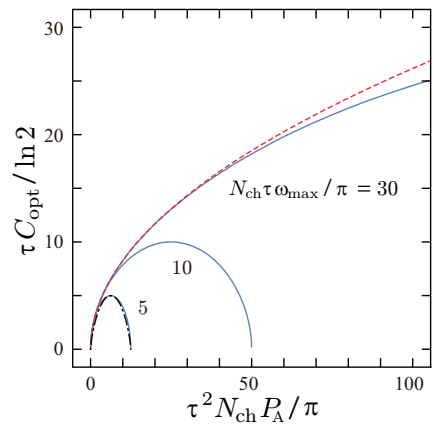}
\caption{
The optimum capacity as a function of signal power $P_A$ for various bandwidths $\omega_{\max}$ (the gap is zero $\omega_{\min}=0$). 
The dot-dashed line indicates the optimum capacity of the narrowband channel, Eq.~(\ref{eqn:cnb}). 
The dashed line indicates the optimum capacity of the wideband channel, Eq.~(\ref{eqn:cwb1}). 
}
\label{fig:opticapa}
\end{figure}

Figure \ref{fig:opticapa} shows the optimum capacity as a function of the signal power of the band-limited channel without the gap $\omega_{\min}=0$. 
The horizontal axis is the heat quantity normalized by $\Delta E$, 
$Q_A/\Delta E=\tau^2 N_{\rm ch} P_A/\pi$. 
%, in Eq.~(\ref{ilft}). 
%
For a small cutoff energy ($j_{\max}=N_{\rm ch} \tau \omega_{\max}/\pi=5$), the curve is well fitted by the optimum capacity of the narrowband channel Eq.~(\ref{eqn:cnb}) indicated by the dot-dashed line. 
The signal power $P_A$ is bounded from above and the maximum is given by Eq.~(\ref{eqn:P_max}). 
The dashed line indicates the optimum capacity of the wideband channel, Eq.~(\ref{eqn:cwb1}). 
With the increase in cutoff energy $\omega_{\max}$, the curve approaches the dashed line.

As we noted, if we change the order in which the zero temperature limit and the analytic continuation are taken, the result mentioned above changes~\cite{YU2015,YU2017}. 
When we set $M=0$ while keeping the inverse temperature $\beta_A$ finite, since $(f_L^\pm)^0 =(\tilde{f}_L^\pm)^0=1$, Eq.~(\ref{eqn:renyi11}) becomes $S_0(\chi)=s_0(\chi)$, see Eq.~(\ref{eqn:bulk_renyi}). 
By taking the limit of zero temperature $\beta_A \to \infty$ while keeping $X=\beta_A \chi$ finite, we obtain, 
%------------------------------------------------------------------------------
\begin{subequations}
\begin{align}
\ln S_0(X)=- \frac{V_A \gamma_A}{2 i X} \, ,  \;\;\;\; (iX<0) \, , 
\label{eqn:renyi0Xv}
\end{align}
%------------------------------------------------------------------------------
where $\gamma_A=C_A \beta_A=\pi^2 \rho_A /3$ is the electronic specific heat coefficient. 
Then the size of the Fock subspace is estimated as, 
%------------------------------------------------------------------------------
\begin{align}
\ln S_{0,Q_A} \approx \sqrt{2 V_A \gamma_A Q_A} \, , 
\label{eqn:renyi0Qv}
\end{align}
\end{subequations}
%------------------------------------------------------------------------------
which may look similar to $\tau C_{\rm opt}$ [Eq.~(\ref{eqn:cwb1})]. 
However, it is not universal and depends on the setup; in order to obtain this form, we assume that the DOS is energy independent in Eq.~(\ref{eqn:bulk_renyi}). 
Moreover, Eq.~(\ref{eqn:renyi0Qv}) depends not on $\tau$ but on $V_A$ and thus is related to bulk states.

%----------------------------------------------------------
%----------------------------------------------------------

\section{Probability distributions}
\label{sec:pdf}

\subsection{Narrowband channel}

For a narrowband channel, $B \ll f$, with perfect transmission, ${\mathcal T}(\omega)=\theta(\omega_{\max}-\omega) \theta(\omega-\omega_{\min})$, the R\'enyi entropy (\ref{eqn:modrenyi0k}) becomes, 
%------------------------------------------------------------------------------
\begin{subequations}
\begin{align}
\ln S_M(X)= \tau 2 N_{\rm ch} B \ln \left[ f_R^+(hf)^M e^{i X h f} + f_R^-(hf)^M \right] \, . \label{eqn:renyi_narrowband}
\end{align}
%------------------------------------------------------------------------------
Then by performing the inverse Fourier transform of Eq.~(\ref{j_renyi}) within the saddlepoint approximation, we obtain, 
%------------------------------------------------------------------------------
\begin{align}
\ln S_M(Q_A) =& \min_{i X \in {\mathbb R}} \left( \ln S_{M}(X) - iX Q_A  \right)
\\ 
=& - M \tau 2 N_{\rm ch} B D(p || q) + (1-M) \nonumber \\ & \times \tau 2 N_{\rm ch}B H_2(P_A/(2 N_{\rm ch} h f B)) \, . 
\label{reny_narrow}
\end{align}
\end{subequations}
%------------------------------------------------------------------------------
Here, 
%------------------------------------------------------------------------------
$D(p \| q) = \sum_j p_j \ln (p_j/q_j)$
%------------------------------------------------------------------------------
is the relative entropy between the distribution $p=( P_A/(2 N_{\rm ch} hfB), 1- P_A/(2 N_{\rm ch} hfB) )$ and $q=( f_R^+(hf),f_R^-(hf) )$, 
which measures the difference between the two distributions $p$ and $q$. 
After the inverse Fourier transform (\ref{eqn:ift1_exact}), we obtain the joint probability distribution, which is the delta distribution;
%------------------------------------------------------------------------------
\begin{align}
P(I_A^\prime,Q_A) =& e^{-\tau 2 N_{\rm ch} B D(p \| q)}
\nonumber \\
&\times \delta(I_A^\prime - \tau 2 N_{\rm ch} B (H_2+D(p \| q))) \, . 
\end{align}
%------------------------------------------------------------------------------
Here $\tau 2 N_{\rm ch} B$ and $\tau P_A/(hf)$ are interpreted respectively as the number of modes and the number of signal quanta, i.e., electrons transmitted to the receiver side. 
When the ratio between these numbers is compatible with the initial electron distribution probability $f_R^+(h f)=P_A/(2 N_{\rm ch} hf B)$, the relative entropy takes its minimum value $D(p \| q)=0$. 
In this case, the transmitted self-information is always $I_A^\prime = \tau 2 N_{\rm ch} B H_2( f_R^+(h f) )$. 

The information-generating function (\ref{c_renyi}) can be derived from Eq.~(\ref{reny_narrow}); 
%------------------------------------------------------------------------------
$\ln S_{1-i \xi,Q_A} = i \xi \, \tau 2 N_{\rm ch} B H_2(P_A/(2 N_{\rm ch} hfB))$. 
%------------------------------------------------------------------------------
Then the conditional self-information is delta distributed as, 
%------------------------------------------------------------------------------
\begin{align}
P_{Q_A}(J) = \delta(J - \tau 2 N_{\rm ch} B  H_2(P_A/(2 N_{\rm ch} hfB)) ) \, .  \label{eqn:pdf_j_nb}
\end{align}
%------------------------------------------------------------------------------
It is independent of the electron distribution probability. 
Therefore the conditional self-information is always, 
%------------------------------------------------------------------------------
\begin{align}
J = \tau 2 N_{\rm ch} B  H_2(P_A/(2 N_{\rm ch} hfB)) \approx \ln \left( \begin{array}{c} \tau 2 N_{\rm ch} B  \\ \tau P_A/(hf) \end{array} \right) \, . 
\label{eqn:J_nb} 
\end{align}
%------------------------------------------------------------------------------
which is the number of possible ways to locate transmitted electrons in available scattering states in subsystem $A$. 
\YU{In order to obtain the last expression in Eq.~(\ref{eqn:J_nb}), we utilized the approximate form of the binomial coefficient, $\ln \left( \begin{array}{c} N \\ n \end{array} \right) \approx  N H_2(n/N) $, which is obtained by applying Stirling's approximation $\ln n! \approx n \ln (n/e)$ for $n \gg 1$. }

\subsection{Wideband channel}

Let us consider the wideband quantum channel, 
$\omega_{\max} \to \infty$, $\omega_{\min} = 0$, and ${\mathcal T}(\omega)=1$. 
The R\'enyi entropy (\ref{eqn:modrenyi0k}) is analytic around $X=0$ and $M=1$: 
%------------------------------------------------------------------------------
\begin{align}
\ln S_M(X) =& \tau P_{\beta} \beta_B^2 \left( \frac{1}{M \beta_B - iX} - \frac{M}{\beta_B} \right) 
\nonumber \\ &
+ \tau \frac{P_\mu}{2} \frac{M \beta_B i X}{M \beta_B-iX} \, , 
\label{eqn:renyiwbc}
\end{align}
%------------------------------------------------------------------------------
where
$P_\mu = N_{\rm ch} g_0^{\rm el} (\mu_B-\mu_A)^2$
is the rate of Joule heat generation and 
$P_{\beta} = N_{\rm ch} g_0 /\beta_B$
is the heat current emitted from subsystem $B$, the right reservoir. 
The coefficients are the conductance quantum $g_0^{\rm el}=1/(2 \pi)$ and the thermal conductance quantum~\cite{RegoRPL1998}, 
$g_0=\pi/6 \times (\beta_A^{-1}+\beta_B^{-1})/2= \pi/ (12 \beta_B)$.

Averages are obtained by performing the derivative Eq.~(\ref{iscum}) as, 
%------------------------------------------------------------------------------
\begin{align}
\langle \! \langle I_A^\prime \rangle \! \rangle = 2 \tau \beta_B P_{\beta} \, ,
\;\;\;\;
\langle \! \langle Q_A \rangle \! \rangle = \tau (P_{\beta}+P_\mu/2) \, .
\label{eqn:avewbc}
\end{align}
%------------------------------------------------------------------------------
When the chemical potential bias is absent, $\mu_A = \mu_B$, we can eliminate $\beta_B$ and obtain, 
%------------------------------------------------------------------------------
\begin{align}
\frac{\langle \! \langle I_A^\prime \rangle \! \rangle}{\tau}
=
\sqrt{
\frac{\pi}{3} N_{\rm ch} \frac{\langle \! \langle Q_A \rangle \! \rangle}{\tau}
}
=C_{\rm WB}(\langle \! \langle Q_A \rangle \! \rangle/\tau)
\, , 
\end{align}
%------------------------------------------------------------------------------
which is the optimum capacity of the wideband channel, Eq.~(\ref{eqn:cwb}). 
The above derivation follows previous approaches in Refs.~\onlinecite{Lebedev1963} and \onlinecite{Pendry1983}. 
The second cumulant, variances and cross correlations, are, 
%------------------------------------------------------------------------------
\begin{subequations}
\begin{align}
\langle \! \langle I_A^{\prime \, 2} \rangle \! \rangle 
=
\beta_B \langle \! \langle I_A^\prime Q_A \rangle \! \rangle 
=
\langle \! \langle I_A^{\prime} \rangle \! \rangle  \, ,
\;\;\;\;
\langle \! \langle Q_A^2 \rangle \! \rangle = 
2 \langle \! \langle Q_A \rangle \! \rangle/\beta_B
\, .
\end{align}
%------------------------------------------------------------------------------
Since the cross correlation is positive, the correlation coefficient is also positive, 
%------------------------------------------------------------------------------
\begin{align}
r = \frac{\langle \! \langle I_A^\prime Q_A \rangle \! \rangle }{\sqrt{\langle \! \langle I_A^{\prime \, 2} \rangle \! \rangle \langle \! \langle Q_A^2 \rangle \! \rangle} } = \sqrt{ \frac{2 P_{\beta}}{2 P_{\beta} + P_\mu} }>0 \, .
\label{eqn:cor_coe}
\end{align}
\end{subequations}
%------------------------------------------------------------------------------
\YU{The correlation coefficient $r$ ranges from -1 to 1 and measures the degree of linear correlation between the two fluctuating variables $I_A^\prime$ and $Q_A$. }
From this relation, we can see that when the chemical potential bias is absent, there is a perfect positive linear correlation $r=1$, which means that there is a one-to-one correspondence between the self-information content and the heat quantity.

\subsubsection{Conditional self-information}

Let us calculate the probability distribution of the conditional self-information. 
First, we perform the inverse Fourier transform of Eq.~(\ref{eqn:renyiwbc}) within the saddlepoint approximation; 
%------------------------------------------------------------------------------
\begin{subequations}
\begin{align}
\ln S_{M}(Q_A) =& \tau C_{\rm WB} \sqrt{ 1+M^2 P_\mu/(2 P_{\beta}) }
\nonumber \\
&- \tau \beta_B M(P_{\beta}+P_\mu/2+P_A) \, .
\label{eqn:reny_qa}
\end{align}
%------------------------------------------------------------------------------
Then, the R\'enyi entropy associated with the probability distribution of the conditional self-information Eq.~(\ref{c_renyi}) becomes, 
%------------------------------------------------------------------------------
\begin{align}
\ln S_{M,Q_A} = \tau C_{\rm WB} \left( \sqrt{ 1+(r^{-2}-1) M^2 } -M r^{-1} \right) \, . 
\label{eqn:mrenyi}
\end{align}
%------------------------------------------------------------------------------
Finally, the probability distribution is obtained by the inverse Fourier transform within the saddlepoint approximation; 
%------------------------------------------------------------------------------
\begin{align}
\ln P_{Q_A}(J)=& \min_{i \xi \in {\mathbb R}} \left( \ln S_{1-i \xi,Q_A} - i \xi J  \right)
\label{eqn:LFTcondprob}
\\
=& \tau C_{\rm WB} \sqrt{ 1 - \frac{ \left[ 1-r J/(\tau C_{\rm WB}) \right]^2}{1-r^2} } - J
\, .
\label{eqn:condprob} 
\end{align}
\end{subequations}
%------------------------------------------------------------------------------

\begin{figure}[h]
\includegraphics[width=0.75 \columnwidth]{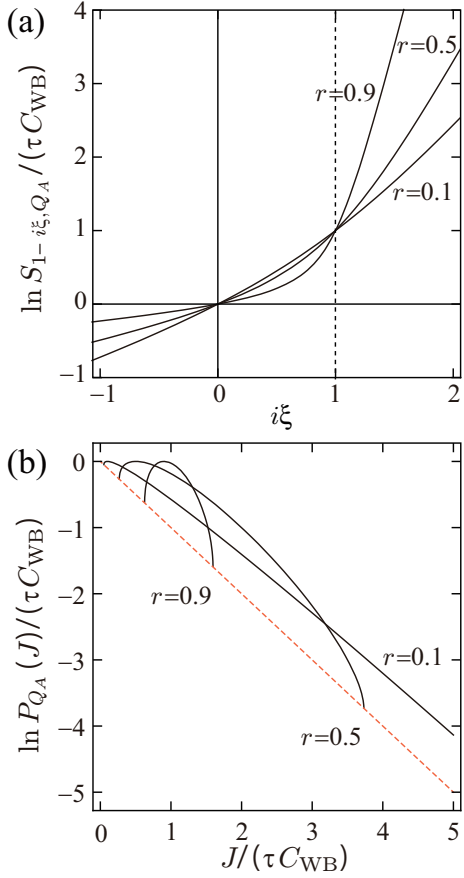}
\caption{
(a) Information-generating function for wideband channel. 
Curves are for a nearly perfectly linearly correlated case ($r=0.9$), for an intermediate case ($r=0.5$), and for a nearly uncorrelated case ($r=0.1$). 
A vertical dotted line indicates the point, $M=0$, where the universal relation Eq.~(\ref{eqn:jeq_wbc}) is satisfied. 
(b) Probability distributions of the conditional self-information content. 
A dashed line corresponds to the exponential distribution Eq.~(\ref{eqn:expdist}). 
}
\label{fig:condself}
\end{figure}

Figure \ref{fig:condself} (a) shows the R\'enyi entropy (\ref{eqn:mrenyi}) for various values of the correlation coefficient. 
We observe that at $M=0$, i.e., $i \xi=1$, all curves intersect. 
Equation (\ref{eqn:mrenyi}) satisfies the universal relation providing the optimum capacity [see Eqs.~(\ref{jareq}) and (\ref{jareq1})] as, 
%------------------------------------------------------------------------------
\begin{align}
\ln S_{0,Q_A} = \tau \, C_{\rm WB} \, . \label{eqn:jeq_wbc}
\end{align}
%------------------------------------------------------------------------------
Panel (b) shows the conditional probability distribution function (\ref{eqn:condprob}). 
The vertical and horizontal axes are normalized by $\tau C_{\rm WB}$. 
The curves are tilted semi-ellipses and depend only on the correlation coefficient $r$. 
The maximum (minimum) is, 
%------------------------------------------------------------------------------
\begin{align}
J_{\max (\min)} =& \lim_{M \to \mp \infty} \frac{\ln S_{M,Q_A} }{1-M}
\nonumber \\
=& \tau \, C_{\rm WB} \left( r^{-1} \pm \sqrt{r^{-2}-1} \right) \, . \label{eqn:maxmin}
\end{align}
%------------------------------------------------------------------------------
Here we utilized the Legendre duality~\cite{Touchette2009} of Eq.~(\ref{eqn:LFTcondprob}) and the fact that a rare event associated with maximum (minimum) $J$ is realized in the limit of $M \to \mp \infty$. 
From Eq.~(\ref{eqn:maxmin}), the width of the distribution is obtained as $J_{\max}-J_{\min}=2 \sqrt{r^{-2}-1}$. 
The width becomes narrower when the two quantities are correlated, as we observe in panel (b). 
For the perfect correlation $r=1$, the delta distribution,
%------------------------------------------------------------------------------
\begin{subequations}
\begin{align}
P_{Q_A}(J) = \delta \left( J - \tau C_{\rm WB} \right) \, , 
\label{eqn:per_lin_cor}
\end{align}
%------------------------------------------------------------------------------
is realized. 
For the uncorrelated case, $r \to 0$, the exponential distribution [dashed line in panel (b)], 
%------------------------------------------------------------------------------
\begin{align}
P_{Q_A}(J) \approx e^{-J} \, , 
\label{eqn:expdist}
\end{align}
\end{subequations}
%------------------------------------------------------------------------------
is approached. 

%----------------------------------------------------------
%----------------------------------------------------------

\subsubsection{Joint probability distribution}

\begin{figure}[h]
\includegraphics[width=0.65 \columnwidth]{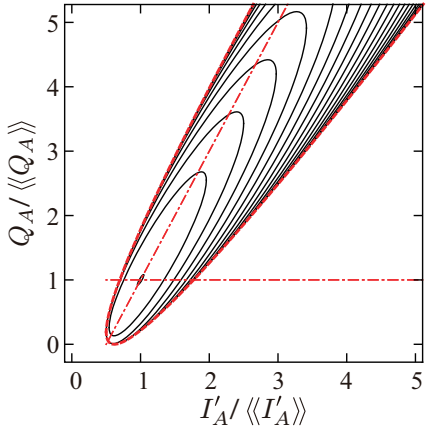}
\caption{
Contour plot of the logarithm of joint probability distribution function of self-information content and heat quantity for wideband channel. 
A thick dotted line indicates the boundary of support Eq.~(\ref{eqn:bousupwb}). 
The correlation coefficient is $r=0.9$. 
The contour interval is $\langle \! \langle I_A^\prime \rangle \! \rangle/4$. 
Two dot-dashed lines indicate Eqs.~(\ref{eqn:bousupwbpc}) and (\ref{eqn:bousupwbuc}). 
}
\label{fig:piqwb}
\end{figure}

The joint probability distribution function is obtained from Eq.~(\ref{eqn:reny_qa}) by applying Eq.~(\ref{eqn:ift1}) as, 
%------------------------------------------------------------------------------
\begin{align}
\ln P(I_A^\prime,Q_A) =&
\min_{M \in {\mathbb R}} \left( \ln S_{M}(Q_A) + M I_A^\prime  \right) -I_A^\prime
\\ =&
\left[ 4 \tau \beta_B P_{\beta} (\tau \beta_B P_{\beta} + \delta I_A^\prime) - 2 P_{\beta}/P_\mu 
\right. \nonumber \\ & \left.\times  
(\beta_B \delta Q_A - \delta I_A^\prime)^2 \right]^{1/2} - I_A^\prime 
\, ,
\label{eqn:wb_jpdf} 
\end{align}
%------------------------------------------------------------------------------
where we introduced 
%------------------------------------------------------------------------------
$\delta I_A^\prime = I_A^\prime - \langle \! \langle I_A^\prime \rangle \! \rangle$ and
$\delta Q_A = Q_A - \langle \! \langle Q_A \rangle \! \rangle$. 
%------------------------------------------------------------------------------
%
Figure \ref{fig:piqwb} is a contour plot of the logarithm of joint probability distribution for $r=0.9$. 
The maximum is $\ln P( \langle \! \langle I_A^\prime \rangle \! \rangle , \langle \! \langle Q_A \rangle \! \rangle )=0$. 
A thick dotted line indicates the boundary of support; 
%------------------------------------------------------------------------------
\begin{align}
\frac{\delta Q_A}{ \langle \! \langle Q_A \rangle \! \rangle} = 2 r^2 \left( \frac{\delta I_A^\prime}{ \langle \! \langle I_A^\prime \rangle \! \rangle} \pm \sqrt{  2 \left( \frac{1}{r^2} -1 \right) \frac{\delta I_A^\prime}{ \langle \! \langle I_A^\prime \rangle \! \rangle} } \right) \, . 
\label{eqn:bousupwb}
\end{align}
%------------------------------------------------------------------------------
The self-information content is bounded from below and the minimum is half of the average self-information, 
%------------------------------------------------------------------------------
\begin{align}
{I_A^\prime}_{\min} = \langle \! \langle I_A^\prime \rangle \! \rangle/2
\, .
\end{align}
%------------------------------------------------------------------------------

As we observe in Eq.~(\ref{eqn:bousupwb}), the width of the distribution vanishes in the perfectly linearly correlated case $r=1$, i.e., $\mu_A = \mu_B$. 
The boundary of support shrinks to,
%------------------------------------------------------------------------------
\begin{subequations}
\begin{align}
\frac{\delta Q_A}{ \langle \! \langle Q_A \rangle \! \rangle} = 2 \frac{\delta I_A^\prime}{ \langle \! \langle I_A^\prime \rangle \! \rangle}
\, , 
\label{eqn:bousupwbpc}
\end{align}
%------------------------------------------------------------------------------
i.e., the fluctuations satisfy $\delta I_A^\prime = \beta_B \delta Q_A$. 
\YU{Here we note that the entropy and average heat Eqs.~(\ref{eqn:avewbc}) satisfy $\langle \! \langle I_A^\prime \rangle \! \rangle = 2 \beta_B \langle \! \langle Q_A \rangle \! \rangle > \beta_B \langle \! \langle Q_A \rangle \! \rangle$, which implies the irreversible nature of the heat transport process~\cite{BlencowePRA2000}. }
For $r \neq 1$, although there is no one-to-one correspondence between the two quantities, we may consider that the two quantities are approximately related as, 
$Q_A/\langle \! \langle Q_A \rangle \! \rangle \propto 2  r^2 I_A^\prime/\langle \! \langle I_A^\prime \rangle \! \rangle $. 
In the limit of uncorrelated case $r \to 0$, which corresponds to $P_{\beta}/P_\mu \to 0$, Eq.~(\ref{eqn:bousupwb}) becomes 
%------------------------------------------------------------------------------
\begin{align}
\frac{\delta Q_A}{ \langle \! \langle Q_A \rangle \! \rangle} \to 0 \,
\label{eqn:bousupwbuc}
\end{align}
\end{subequations}
%------------------------------------------------------------------------------
In Fig.~\ref{fig:piqwb}, Eqs.~(\ref{eqn:bousupwbpc}) and (\ref{eqn:bousupwbuc}) are indicated by dot-dashed lines.

%----------------------------------------------------------
%----------------------------------------------------------

\YU{
\subsection{Short summary of Secs.~\ref{sec:opt_cap} and \ref{sec:pdf}}
\label{sec:i_summary}

In Secs.~\ref{sec:opt_cap} and \ref{sec:pdf}, we provided rather detailed derivations. 
Here, we summarize relevant results in these two sections. 

\begin{itemize}

\item

The R\'enyi entropy of order zero is related to the generating function of integer partitions; 
%------------------------------------------------------------------------------
\begin{align}
S_0(X=\lambda/\Delta E) \approx \prod_{j \in {\mathcal S}} \left( 1+e^{i \lambda j} \right) \, , 
\tag{\ref{gefuip}}
\end{align}
%------------------------------------------------------------------------------
where ${\mathcal S}=\{ j_{\min}, \cdots, j_{\max} \}$ and $j= \omega/\Delta E$ is assumed to be integers. 
The energy resolution $\Delta E=h/(2 N_{\rm ch} \tau)$ is due to the energy-time uncertainty relation. 
The expression is independent of details of the mesoscopic quantum electric conductor and only depends on the statistics of particles and the bandwidth. 

The optimum capacity for the narrowband case is, 
%------------------------------------------------------------------------------
\begin{align}
C_{\rm NB}(P_A) = 2 N_{\rm ch} B H_2(P_A/(2 N_{\rm ch} h f B)) \, , 
\tag{\ref{eqn:cnb}}
\end{align}
%------------------------------------------------------------------------------
where $H_2(x)=-x \ln x -(1-x) \ln (1-x)$ is the binary entropy. 
Here, $\tau 2 N_{\rm ch} B$ and $\tau P_A/(hf)$ are the number of modes and the number of signal quanta. 
Then $e^{\tau C_{\rm NB}}$ is regarded as the number of possible ways to distribute signal quanta into available modes. 

The optimum capacity for the wideband case is, 
%------------------------------------------------------------------------------
\begin{align}
C_{\rm WB} (P_A) = \frac{\pi}{\tau} \sqrt{\frac{Q_A/\Delta E}{3}} \, . \tag{\ref{eqn:cwb1}}
\end{align}
%------------------------------------------------------------------------------
The above mentioned results derived systematically from the R\'enyi entropy (\ref{eqn:modrenyi0k}) based on the microscopic Hamiltonian, reproduce previous theories, see Refs.~\onlinecite{CavesRMP1994} and \onlinecite{BlencowePRA2000}. 

\item
The conditional self-information for the narrowband case is delta distributed as, 
%------------------------------------------------------------------------------
\begin{align}
P_{Q_A}(J) = \delta(J - \tau \, C_{\rm NB}(P_A) )
\, .  \tag{\ref{eqn:pdf_j_nb}}
\end{align}
%------------------------------------------------------------------------------
Thus, the conditional self-information does not fluctuate. 

For the wideband channel, 
%------------------------------------------------------------------------------
\begin{align}
\ln P_{Q_A}(J)= \tau C_{\rm WB} \sqrt{ 1 - \frac{ \left[ 1-r J/(\tau C_{\rm WB}) \right]^2}{1-r^2} } - J
\, , 
\tag{\ref{eqn:condprob}}
\end{align}
%------------------------------------------------------------------------------
which depends on the correlation coefficient, 
%------------------------------------------------------------------------------
\begin{align}
r = \frac{\langle \! \langle I_A^\prime Q_A \rangle \! \rangle }{\sqrt{\langle \! \langle I_A^{\prime \, 2} \rangle \! \rangle \langle \! \langle Q_A^2 \rangle \! \rangle} } 
= \sqrt{ \frac{ \langle \! \langle I_A^\prime \rangle \! \rangle }{2 \beta_B \langle \! \langle Q_A \rangle \! \rangle } } \, . 
\tag{\ref{eqn:cor_coe}}
\end{align}
%------------------------------------------------------------------------------
It measures how much two variables, $I_A^\prime$ and $Q_A$, are linearly correlated and satisfies $0 \leq r \leq 1$. 
For $r = 1$, the two variables are perfectly linearly correlated and there is one-to-one correspondence between the two quantities. 
It is realized when the chemical potential bias is absent, $\mu_A=\mu_B$, and the averages of self-information and heat quantity satisfy, 
$\langle \! \langle I_A^\prime \rangle \! \rangle = 2 \beta_B \langle \! \langle Q_A \rangle \! \rangle = \tau N_{\rm ch} \pi/(6 \beta_{\rm B})$. 
In this case, Eq.~(\ref{eqn:condprob}) is reduced to the delta distribution (\ref{eqn:per_lin_cor}), 
%------------------------------------------------------------------------------
$P_{Q_A}(J) = \delta \left( J - \tau C_{\rm WB} \right)$. 
%------------------------------------------------------------------------------
When the chemical potential bias is much larger than the temperature bias 
$|\mu_B - \mu_A| \gg \beta_B^{-1}$, the two quantities become uncorrelated, $r \to 0$. 
In this case, Eq.~(\ref{eqn:condprob}) becomes the exponential distribution 
%------------------------------------------------------------------------------
$P_{Q_A}(J) \approx e^{-J}$, Eq.~(\ref{eqn:expdist}).
%------------------------------------------------------------------------------

\end{itemize}
}

\section{resonant tunneling and Coulomb interaction}
\label{sec:rtcb}

\subsection{Energy-dependent transmission probability}

\begin{figure}[h]
\includegraphics[width=0.65 \columnwidth]{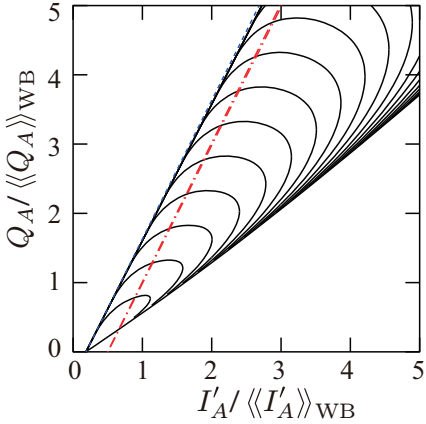}
\caption{
Contour plot of the logarithm of joint probability distribution function of self-information content and heat quantity close to the resonant tunneling condition. 
The contour interval is $\langle \! \langle I_A^\prime \rangle \! \rangle_{\rm WB}/4$. 
A dotted line indicates Eq.~(\ref{eqn:rlmlb}). 
A dot-dashed line is the boundary of support for the wideband channel Eq.~(\ref{eqn:bousupwb}). 
Axes are normalized by the corresponding values of the wideband channel $\langle \! \langle I_A^\prime \rangle \! \rangle_{\rm WB}$ and $\langle \! \langle Q_A \rangle \! \rangle_{\rm WB}$ [Eqs.~(\ref{eqn:avewbc})]. 
The average values are 
$\langle \! \langle I_A^\prime \rangle \! \rangle = 0.371 \langle \! \langle I_A^\prime \rangle \! \rangle_{\rm WB}$ and 
$\langle \! \langle Q_A \rangle \! \rangle = 0.139 \langle \! \langle Q_A \rangle \! \rangle_{\rm WB}$. 
Parameters: 
$\beta_R \Gamma=1$, $\mu_R=0.01 \Gamma$ and $\omega_{\rm max}=10^3 \Gamma$. 
}
\label{fig:piqrlm}
\end{figure}

In this section, we consider the resonant tunneling condition $\Gamma_L=\Gamma_R$ and $\mu_A = \epsilon_D = 0$, where the transmission probability is
%------------------------------------------------------------------------------
\begin{align}
{\mathcal T}(\omega) = \frac{1}{1+4 (\omega/\Gamma)^2}
\, . 
\end{align}
%------------------------------------------------------------------------------
Figure~\ref{fig:piqrlm} is a contour plot of the logarithm of joint probability distribution of self-information content and heat quantity obtained by numerically solving Eqs.~(\ref{eqn:doubleLFT}) and (\ref{eqn:modrenyi0k}). 
In this panel, the voltage difference is small $\mu_B=0.01 \Gamma$ and the temperature of the subsystem $B$ is comparable to the level broadening $\beta_B \Gamma=1$. 
A dot-dashed line indicates the boundary of support for the wideband channel (\ref{eqn:bousupwb}), i.e., the result when ${\mathcal T}=1$, which implies a perfect linear correlation, $I_A^\prime \approx \beta_B Q_A$ [Eq.~(\ref{eqn:bousupwbpc})]. 
We checked that the perfect linear correlation is approached when the temperature is low $\beta_B \Gamma \ll1$. 
In Fig.~\ref{fig:piqrlm}, since the temperature is comparable to the level broadening, the perfect linear correlation is spoiled. 

A dotted line indicates the minimum self-information content for a given heat content. 
It is almost parallel to the dot-dashed line. 
The minimum can be estimated in the following. 
For $M \to \infty$, the R\'enyi entropy (\ref{eqn:modrenyi0k}) is approximately, 
%------------------------------------------------------------------------------
\begin{subequations}
\begin{align}
\ln S_M(X) \approx & M I_{\rm m} + \tau \frac{\pi}{12} \frac{1}{i X - \beta_B M}\, , \;\;\;\; (i X < \beta_B M) \, , 
\\
I_{\rm m} = & \frac{\tau}{\pi} \int_0^\infty d \omega \ln \left( 1- {\mathcal T}(\omega) f_R^+(\omega) \right) \, . 
\end{align}
%------------------------------------------------------------------------------
Then after a few steps of calculations, the minimum is obtained as, 
%------------------------------------------------------------------------------
\begin{align}
{I_A^\prime}_{\min}= I_{\rm m} + \tau \beta_B P_A \, , 
\label{eqn:rlmlb}
\end{align}
\end{subequations}
%------------------------------------------------------------------------------
where we used the Legendre duality 
%------------------------------------------------------------------------------
${I_A^\prime}_{\min}=\lim_{i \xi \to -\infty} \ln S_{1- i \xi} (Q_A)/(i \xi)$. 
%------------------------------------------------------------------------------

%----------------------------------------------------------
%----------------------------------------------------------

\begin{figure}[h]
\includegraphics[width=0.7 \columnwidth]{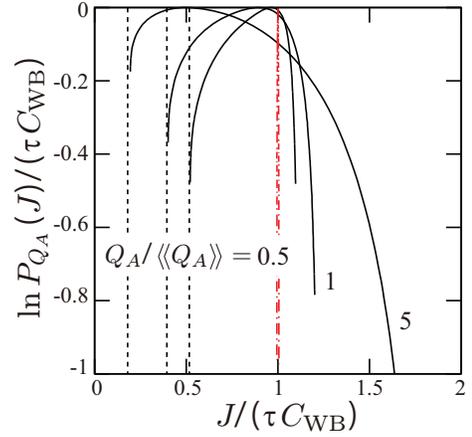}
\caption{
Probability distribution of conditional self-information close to the resonant tunneling condition. 
Curves are for various values of heat quantity, 
$Q_A/\langle \! \langle Q_A \rangle \! \rangle=0.5,1$ and 5. 
A dot-dashed line indicates the result of the wideband channel (\ref{eqn:condprob}). 
Dotted lines indicate the minimum of conditional self-information Eq.~(\ref{eqn:rlmlbcp}). 
The parameters are the same as those in Fig.~\ref{fig:piqrlm}. 
}
\label{fig:pjcond}
\end{figure}

Figure~\ref{fig:pjcond} shows the probability distribution of conditional self-information for various values of heat quantity. 
A dot-dashed line indicates the result of the wideband channel (\ref{eqn:condprob}). 
The vertical and horizontal axes are normalized by the optimum capacity of the wideband channel. 
With an increase in signal power, the peak position shifts leftward, which means that the transmitted information decreases as compared with that of the wideband channel. 
At the same time, the width increases, which means that the number of typical sequences decreases. 
Dotted lines indicate the minimum of conditional self-information corresponding to the dotted line in Fig.~\ref{fig:piqrlm}. 
The minimum in Fig.~\ref{fig:piqrlm} and that in Fig.~\ref{fig:pjcond} are related as we can deduce from Eq.~(\ref{cond_and_joint_pdf}); 
%------------------------------------------------------------------------------
\begin{align}
J_{\min}={I_A^\prime}_{\min} + \ln P(S_A) \, .
\label{eqn:rlmlbcp}
\end{align}
%------------------------------------------------------------------------------

%----------------------------------------------------------
%----------------------------------------------------------

\subsection{Coulomb interaction }
\label{sec:couint}

Here we discuss the effect of the on-site Coulomb interaction. 
For this purpose, we adopt the same model in Ref.~\onlinecite{YU2015}. 
Namely, we introduce the spin degree of freedom, $\hat{a}_{rk} \to \hat{a}_{rk \sigma}$ and $\hat{d} \to \hat{d}_{\sigma}$ ($\sigma=\uparrow,\downarrow$). 
The on-site Coulomb interaction is included in the Hamiltonian of subsystem $B$; 
%------------------------------------------------------------------------------
\begin{align}
\hat{H}_B =
\sum_{k \sigma} \epsilon_{Rk} \hat{a}_{R k \sigma}^\dagger \hat{a}_{R k \sigma} + \sum_{\sigma} \epsilon_D \hat{d}_\sigma^\dagger \hat{d}_\sigma + U \hat{d}_\uparrow^\dagger \hat{d}_\uparrow \hat{d}_\downarrow^\dagger \hat{d}_\downarrow 
\, . 
\end{align}
%------------------------------------------------------------------------------
Because of the spin degree of freedom, the number of channels is doubled $N_{\rm ch} =2$. 

\begin{figure}[h]
\includegraphics[width=0.9 \columnwidth]{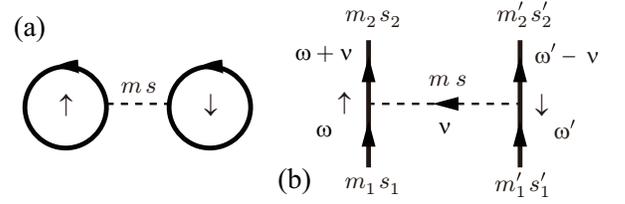}
\caption{
Diagrams correspond to (a) the Hartree term and (b) the bare interaction vertex. 
Thick solid lines correspond to the full Green function matrix, Eq.~(\ref{eqn:fgfm}). 
Dotted lines indicate the Coulomb interaction. 
}
\label{fig:feyndiag}
\end{figure}

We perform the perturbative expansion of the Keldysh partition function~\cite{YU2015} (\ref{eqn:keldysh_partition_function}) in powers of the Coulomb interaction $U$. 
The zeroth-order contribution is Eq.~(\ref{eqn:modrenyi0k}). 
The Hartree term, the first-order contribution, is depicted in Fig.~\ref{fig:feyndiag} (a); 
%------------------------------------------------------------------------------
\begin{align}
&
i U \sum_{m=1}^M \sum_{s= \pm} s \int_0^\tau dt G_{D}^{ \{ \chi_j \} } (t_{ms},t_{ms}) G_{D}^{ \{ \chi_j \} }  (t_{ms},t_{ms}) \nonumber \\
=& i U M \tau \sum_{s= \pm} s \int_0^\tau dt \left( G_{D}^{ \{ \chi_j \} }  (t_{ms},t_{ms})
\right)^2 \nonumber \\
=& i U M \tau \sum_{s= \pm} s \left( \int \frac{d \omega}{2 \pi} \left[ {\bm U}^\dagger {\bm G}_{D}^{\bar{\chi}} (\omega) {\bm U} \right]_{ms,ms} \right)^2 \nonumber \\ =& \tau 2 U M n_{M,\bar{\chi},q} \delta n_{M,\bar{\chi}} \, . 
\end{align}
%------------------------------------------------------------------------------
Since $\left[{\bm U}^\dagger {\bm G}_{D}^{\bar{\chi}} {\bm U} \right]_{ms,ms} = \left[{\bm G}_{D}^{\bar{\chi}} \right]_{ms,ms} $, the result is independent of the phase $\phi_m$. 
The classical and quantum components of electron occupancy inside the dot are calculated by using the local Green function matrix~(\ref{eqn:rlgfm})
%------------------------------------------------------------------------------
\begin{align}
\delta n_{M,\bar{\chi}} =& \int \frac{d \omega}{2 \pi} \frac{G_{D}^{\bar{\chi},m+,m+}(\omega) + G_{D}^{\bar{\chi},m-,m-} (\omega)}{2 i} \nonumber \\
=& \int_{-\omega_{\max}}^{\omega_{\max}} d \omega \left( 1-\frac{1}{M} \frac{\partial_{\epsilon_D} \ln \Omega_{M,-\bar{\chi}S_A(\omega)}(\omega)}{\partial_{\epsilon_D} \ln \rho(\omega)} \right)
\nonumber \\ & \times \rho(\omega) \sum_r \frac{\Gamma_r}{\Gamma} (f_r^+(\omega)-1/2) \, , \\
n_{M,\bar{\chi},q} =& \int_{-\omega_{\max}}^{\omega_{\max}} \frac{d \omega}{2 \pi} (G_{D}^{\bar{\chi},m-,m-}(\omega) - G_{D}^{\bar{\chi},m+,m+} (\omega)) \nonumber \\
=& \partial_{\epsilon_D} \ln S_M(\bar{\chi})/(M N_{\rm ch} \tau)
\, , 
\end{align}
%------------------------------------------------------------------------------
where we introduced the cutoff energy $\omega_{\max}$. 
Since $ \delta n_{0,\bar{\chi}}$ and $ n_{0,\bar{\chi},q}$ are finite, we confirm that the Hartree term vanishes when we take the limit $M \to 0$. 
Therefore, the optimum capacity is not affected by the weak Coulomb interaction.

As we mentioned, the first-order contribution is independent of the phase $\phi_m$. 
The same is true for any closed diagram. 
Let us analyze the interaction vertex on the $s$ branch of the $m$th Keldysh contour, Fig.~\ref{fig:feyndiag} (b); 
%------------------------------------------------------------------------------
\begin{align}
&
s U \left[{\bm U}^\dagger {\bm G}_{D}^{\bar{\chi}}(\omega+\nu) {\bm U} \right]_{m_2 s_2,m s} \left[{\bm U}^\dagger {\bm G}_{D}^{\bar{\chi}}(\omega) {\bm U} \right]_{m s,m_1 s_1}
\nonumber \\
\times &
\left[{\bm U}^\dagger {\bm G}_{D}^{\bar{\chi}}(\omega^\prime-\nu) {\bm U} \right]_{m_2^\prime s_2^\prime,m s} \left[{\bm U}^\dagger {\bm G}_{D}^{\bar{\chi}}(\omega^\prime) {\bm U} \right]_{m s,m_1^\prime s_1^\prime}
\nonumber \\
=&
s U G_{D}^{ \bar{\chi} , m_2 s_2,m s}(\omega+\nu) G_{D}^{ \bar{\chi} , m s,m_1 s_1}(\omega) 
\nonumber \\
\times &
G_{D}^{ \bar{\chi} , m_2' s_2',m s}(\omega'-\nu) G_{D}^{ \bar{\chi} , m s,m_1' s_1'}(\omega') 
\nonumber \\
& \times
e^{i [\phi_{m_2}(\omega+\nu)+\phi_{m_2'}(\omega'-\nu)-\phi_{m_1}(\omega)-\phi_{m_1'}(\omega')]}
\, . 
\end{align}
%------------------------------------------------------------------------------
It is independent of phase $\phi_m$ defined on the $m$th Keldysh contour. 
The phase cancels because of the conservation of energy;  
$-\phi_m(\omega+\nu)-\phi_m(\omega'-\nu)+\phi_m(\omega)+\phi_m(\omega') =0$. 
Therefore, any closed diagram is independent of the phase $\phi_m$, since at each bare vertex, the phase cancels~\cite{SU2008}. 

The above discussion implies that at the steady state, operators $\hat{\rho}_A$ and $\hat{H}_A$ are effectively commutative even in the presence of the intra-Coulomb interaction. 
This is because the energy associated with the coupling between the two subsystems $\hat{V}$ is negligible compared to the net energy transferred to subsystem $A$, which grows linearly in $\tau$. 
Because $\hat{\rho}_A$ and $\hat{N}_A$ are commutative~\cite{YU2017}, we expect, 
%------------------------------------------------------------------------------
\begin{align}
[\hat{\rho}_A(\tau) , \hat{Q}_A] \approx 0
\, ,
\label{eqn:lesc}
\end{align}
%------------------------------------------------------------------------------
in the steady state. 
In other words, in the steady state, the {\it local} heat quantity is a classical quantity, as anticipated.

%----------------------------------------------------------
%----------------------------------------------------------

\section{previous approach}
\label{sec:previous_theories}

We compare our approach and the previous quantum information theory approach~\cite{CavesRMP1994,BlencowePRA2000}. 
The previous approach is as follows. 
The communication channel is characterized by an input (output) alphabet $B$ ($A$) with letters labeled $b$ ($a$). 
The input letter $b$ is encoded in a quantum state $\hat{\rho}_b$. 
The probability of transmitting the input letter $b$ is $p_B (b) $. 
The conditional probability of output letter $a$ given input letter $b$ is 
%------------------------------------------------------------------------------
$p_{A|B}(a|b) = {\rm Tr} \hat{\rho}_b \hat{F}_a$, 
%------------------------------------------------------------------------------
where $\hat{F}_a$ is the effect satisfying $\sum_a \hat{F}_a = \hat{1}$. 
The capacity is the average mutual information $H(A;B)$ maximized over all possible input distributions $p_B(b)$, 
%------------------------------------------------------------------------------
\begin{subequations}
\begin{align}
C &=\frac{1}{\tau} \max_{ \{ p_B(b) \} } H(A;B) \, , 
\nonumber \\
H(A;B) &= \sum_b p_B(b) \sum_a p_{A|B}(a|b) \ln \frac{p_{A|B}(a|b)}{p_A(a)} \, , 
\end{align}
%------------------------------------------------------------------------------
where $p_A(a)=\sum_b p_{A|B}(a|b) p_B(b)$ is the probability to obtain the output $a$. 
A further maximization over measurement schemes and over input states yields the optimum capacity; 
$C_{\rm opt}= \max_{ \{ \hat{\rho}_b \} } \max_{ \{ \hat{F}_a \} } C$. 
%------------------------------------------------------------------------------
By exploiting Holevo's theorem, 
%------------------------------------------------------------------------------
$\max_{ \{ \hat{F}_a \} } H(A;B) \leq S( \hat{\rho} ) - \sum_b p_B(b) S(\hat{\rho}_b)$, 
%------------------------------------------------------------------------------
where $\hat{\rho} = \sum_b p_B(b) \hat{\rho}_b$, one can find a link between the mutual information and the von-Neumann entropy: 
%------------------------------------------------------------------------------
$\max_{ \{ \hat{\rho}_b \} } \max_{ \{ \hat{F}_a \} } \max_{ \{ p_B(b) \} } H(A;B) \leq \max_{ \hat{\rho} } S( \hat{\rho} )$. 
%------------------------------------------------------------------------------
The maximum turned out to be the optimum capacity; 
%------------------------------------------------------------------------------
\begin{align}
\tau C_{\rm opt} = \max_{\hat{\rho}} S(\hat{\rho}) = \ln {\rm rank} \hat{\rho} \, . \label{optc_caves}
\end{align}
\end{subequations} 
%------------------------------------------------------------------------------
\YU{
The rank of the density matrix $\hat{\rho}$ is estimated by counting the number of possible particle-number eigenstates~\cite{CavesRMP1994,BlencowePRA2000}. 
In the following, we assume only electrons above the Fermi energy carry the information. 
For a linear dispersion channel, allowed energies are $\Delta E \, j$, 
where $j \in {\mathcal S} = \{ 1,2,\cdots \}$ and $\Delta E=h/\tau$ is the minimum level spacing. 
Then the rank of $\hat{\rho}$ is the number of Fock states, 
%------------------------------------------------------------------------------
\begin{subequations}
\begin{align}
| n_1,n_2, \cdots \rangle = | \{ n_j \} \rangle \, , 
\end{align}
%------------------------------------------------------------------------------
where $n_j=0,1$ is the electron occupation number of the mode $j$. 
The signal energy corresponds to the energy of the Fock state $| \{ n_j \} \rangle$ as, 
%------------------------------------------------------------------------------
\begin{align}
P \tau = \sum_{j=1}^\infty \Delta E j \, n_j \, . 
\end{align}
%------------------------------------------------------------------------------
Therefore, when 
%------------------------------------------------------------------------------
$P \tau /\Delta E = \sum_{j=1}^\infty j \, n_j$
%------------------------------------------------------------------------------
is a positive integer, the number of Fock states with a given energy $P \tau$ is the number of integer partitions into distinct elements of the set ${\mathcal S}$, $p(P \tau /\Delta E| {\rm distinct \, parts \, in } \, {\mathcal S})$. 
The partition function is, for example, $p(6| {\rm distinct \, parts \, in } \, {\mathcal S})=4$ since $6$ can be partitioned into 4 ways $6=1+5=2+4=1+2+3$. 
In the end, we obtain, 
%------------------------------------------------------------------------------
\begin{align}
\tau C_{\rm opt} = \ln  p(P \tau /\Delta E| {\rm distinct \, parts \, in } \, {\mathcal S}) \, . \label{optc_caves_fermions}
\end{align}
\end{subequations}
%------------------------------------------------------------------------------
which is the result obtained previously for fermions in Ref.~\onlinecite{BlencowePRA2000}. 
The above mentioned derivation was first applied to bosons in Ref.~\onlinecite{CavesRMP1994}. 
}

One may think that Eq.~(\ref{optc_caves}) is equivalent to Eq.~(\ref{eqn:2ndlaw}), if one regards $\hat{\rho}$ here as $\hat{\rho}_{A,Q_A}$. 
Precisely speaking, we consider that Eqs.~(\ref{eqn:2ndlaw}) and (\ref{optc_caves}) would be different. 
In Ref.~\onlinecite{BlencowePRA2000}, it was pointed out that the operators $\hat{F}_a$ and $\hat{\rho}_b$ act on the Fock subspace of left-moving states (in our setup, the information flows from right to left; see Fig.~\ref{setup}). 
Thus, the RHS of Eq.~(\ref{optc_caves}) is the logarithm of the size of the Fock subspace of left-movers containing a given total energy. 
In our approach, the operator $\hat{\rho}_{A,Q_A}$ acts {\it locally} on the subsystem $A$, the receiver side. 
Therefore, our approach accounts for the spatial separation between the transmitter side and the receiver side to a certain extent. 
On the other hand, we did not calculate the mutual information. 
Indeed, we do not know how to calculate it based on the Keldysh technique. 
This problem is beyond the scope of the present paper.

\section{summary}
\label{sec:summary}

In summary, we have investigated fluctuations of self-information and heat quantity. 
We bipartition the quantum conductor and regard subsystem $A$ ($B$) as the receiver (transmitter) side and considered the reduced-density matrix of subsystem $A$. 
By exploiting the multi-contour Keldysh Green function technique, we calculate the R\'enyi entropy of a positive integer order subjected to the constraint of the local heat quantity of subsystem $A$. 
By performing the analytic continuation, we relate it to the information-generating function. 
When the thermal noise of the receiver side is absent, there exists the Jarzynski equality-like universal relation Eq.~(\ref{jareq}), which relates the R\'enyi entropy of order 0 at the steady state with the optimum capacity of information transmission. 
For electrons, the optimum capacity is related to the number of integer partitions into distinct parts. 
The optimum capacity obtained in this way is consistent with that of the quantum information theory approach~\cite{CavesRMP1994,BlencowePRA2000}. 

We applied our theory to the resonant-level model. 
The expressions of average self-information and average heat quantity are consistent with those of the previous scattering theory~\cite{Sivan1986,Akkermans2009}. 
We analyzed the fluctuations of self-information and conditional self-information for a narrowband channel, for a wideband channel, and for a resonant tunneling condition. 
We calculated the correction to the R\'enyi entropy induced by the on-site Coulomb interaction within the Hartree approximation and checked that the weak Coulomb interaction does not alter the optimum capacity. 

We also pointed out that in the steady state, even in the presence of the intra-Coulomb interaction, the reduced-density matrix of subsystem $A$ may be diagonal in the eigenstates of the operator of ``local heat quantity" acting locally on subsystem $A$.

We thank Hiroki Okada and Yasuhiro Tokura for the valuable discussions. 
This work was supported by JSPS KAKENHI Grants 17K05575 and JP26220711.

\appendix

\section{Projection operator}
\label{projection_operator}

\YU{
Here we relate Eq.~(\ref{proj_sa}) with the standard form of the projection operator~\cite{Wiseman_Milburn}. 
Let $| S_A,j \rangle$ be an orthonormal basis such that, 
%------------------------------------------------------------------------------
\begin{align}
\hat{S}_{A} |S_A,j \rangle = S_A |S_A,j \rangle \, , \;\;\;\; j \in \{1, \cdots, N_{S_A} \} \, . 
\end{align}
%------------------------------------------------------------------------------
Here the index $j$ is used to label possible degeneracies. 
We assumed that the dimensionless heat quantity is discrete, $S_A=\Delta n$, where $n$ is an integer. 
Then, we obtain, 
%------------------------------------------------------------------------------
\begin{align}
\langle {S_A}',j' | \hat{\Pi}_{{S_A}} | {S_A}'',j'' \rangle =& \frac{\Delta}{2 \pi} \int_{-\pi/\Delta}^{\pi/\Delta} d \chi e^{-i \chi ({S_A}-{S_A}')} \nonumber \\ & \times \delta_{{S_A}',{S_A}''} \, \delta_{j',j''} \nonumber \\ =& \delta_{{S_A},{S_A}'} \delta_{{S_A}',{S_A}''} \delta_{j',j''} \, .
\end{align}
%------------------------------------------------------------------------------
By combining it with the completeness relation, 
%------------------------------------------------------------------------------
\begin{align}
\sum_{S_A} \sum_{j=1}^{N_{S_A}} |S_A,j \rangle \langle S_A,j | = \hat{1} \, .
\end{align}
%------------------------------------------------------------------------------
the projection operator is rewritten as, 
%------------------------------------------------------------------------------
\begin{align}
\hat{\Pi}_{{S_A}} = \sum_{j=1}^{N_{{S_A}}} |{S_A},j \rangle \langle {S_A},j |
\, , \label{eqn:rand_n_projector}
\end{align}
%------------------------------------------------------------------------------
which is the standard form of the rank-$N_{S_A}$ projector (see Chap. 1.2.2 of Ref.~\onlinecite{Wiseman_Milburn}). 
From Eq.~(\ref{eqn:rand_n_projector}), one can derive, 
%------------------------------------------------------------------------------
\begin{align}
\hat{\Pi}_{{S_A}} \hat{\Pi}_{{S_A}'} = \delta_{S_A,{S_A}'} \, \hat{\Pi}_{S_A} 
\, .
\end{align}
%------------------------------------------------------------------------------
}

\YU{
\section{Derivation of Eq.~(\ref{eqn:noncommutativity})}
\label{derivation_non_commutativity}

Here we write the initial density matrix as $\hat{\rho}_{{\rm eq}} = \hat{\rho}_{A {\rm eq}} \, \hat{\rho}_{B {\rm eq}}$. 
Our setup satisfies the following conditions. 

\noindent
(i) The {\it total} particle number is conserved during the time evolution, 
%------------------------------------------------------------------------------
\begin{align}
[ \hat{H}, \hat{N}_{A}+\hat{N}_{B} ] = 0 \, . \label{eqn:particle_conservation}
\end{align}
%------------------------------------------------------------------------------
\noindent
(ii) The initial state is diagonal in the particle number sector, 
%------------------------------------------------------------------------------
\begin{align}
[ \hat{\rho}_{{\rm eq}}, \hat{N}_{A}+\hat{N}_{B} ] = 0 \, , \label{eqn:particle_sector}
\end{align}
%------------------------------------------------------------------------------
and in the energy sector of the unperturbed Hamiltonian, 
%------------------------------------------------------------------------------
\begin{align}
[ \hat{\rho}_{{\rm eq}}, \hat{H}_{A}+\hat{H}_{B} ] = 0 \, . \label{eqn:energy_sector}
\end{align}
%------------------------------------------------------------------------------

Then, the commutation relation, the LHS of Eq.~(\ref{eqn:noncommutativity}), is, 
%------------------------------------------------------------------------------
\begin{align}
[\hat{\rho}_{A}(\tau), \hat{Q}_{A}]  = [ \hat{\rho}_{A}(\tau), \hat{H}_A ] - \mu_A [ \hat{\rho}_{A}(\tau), \hat{N}_A ] \,. \label{eqn:rho_q}
\end{align}
%------------------------------------------------------------------------------
The second term on the RHS of Eq.~(\ref{eqn:rho_q}) is further calculated as, 
%------------------------------------------------------------------------------
\begin{align}
[ \hat{\rho}_A(\tau), \hat{N}_A ] =& {\rm Tr}_B \left( [ e^{-i \hat{H} \tau} \hat{\rho}_{{\rm eq}} e^{i \hat{H} \tau}, \hat{N}_A + \hat{N}_B ] \right) \nonumber \\ & - {\rm Tr}_B \left( [ e^{-i \hat{H} \tau} \hat{\rho}_{{\rm eq}} e^{i \hat{H} \tau}, \hat{N}_B ] \right) \, . \label{eqn:local_particle_number_ss}
\end{align}
%------------------------------------------------------------------------------
The first line of the RHS is zero because of Eqs.~(\ref{eqn:particle_conservation}) and (\ref{eqn:particle_sector}). 
The second line of the RHS is also zero from the cyclic property of the partial trace over the subsystem $B$, 
%------------------------------------------------------------------------------
\begin{align}
{\rm Tr}_B \left( [ \hat{\cal O} , \hat{N}_B ] \right) = 0 \, . \label{eqn:cyclic_property}
\end{align}
%------------------------------------------------------------------------------
Here, an operator $\hat{\cal O}$ acts on the subsystems $A$ and $B$. 
Therefore, Eq.~(\ref{eqn:local_particle_number_ss}) is zero, which is the consequence of the local-particle number super-selection (see Appendix A of Ref.~\onlinecite{YU2017}). 

By exploiting Eqs.~(\ref{eqn:energy_sector}) and (\ref{eqn:cyclic_property}), the first term on the RHS of Eq.~(\ref{eqn:rho_q}) is transformed as, 
%------------------------------------------------------------------------------
\begin{align}
[ \hat{\rho}_A(\tau), \hat{H}_A ] =& {\rm Tr}_B \left( [ e^{-i \hat{H} \tau} \hat{\rho}_{{\rm eq}} e^{i \hat{H} \tau} , \hat{H} ] \right) \nonumber \\ &- {\rm Tr}_B \left( [ e^{-i \hat{H} \tau} \hat{\rho}_{{\rm eq}} e^{i \hat{H} \tau}, \hat{H}_B + \hat{V} ] \right) \\ =& {\rm Tr}_B \left( e^{-i \hat{H} \tau} [  \hat{\rho}_{{\rm eq}} , \hat{V} ] e^{i \hat{H} \tau} \right) \nonumber \\ &-{\rm Tr}_B \left( [ e^{-i \hat{H} \tau} \hat{\rho}_{{\rm eq}} e^{i \hat{H} \tau}, \hat{V} ] \right) \\ =& {\rm Tr}_B \left( e^{-i \hat{H} \tau} \, [  \hat{\rho}_{{\rm eq}} , \hat{V} - e^{i \hat{H} \tau} \hat{V} e^{-i \hat{H} \tau} ] \, e^{i \hat{H} \tau} \right) \, . \label{eqn:absence_local_energy_ss}
\end{align}
%------------------------------------------------------------------------------
In general, Eq.~(\ref{eqn:absence_local_energy_ss}) is not necessarily zero. 
By summarizing above, we obtain Eq.~(\ref{eqn:noncommutativity}). 
}

\section{Explicit expressions of the multi-contour Keldysh Green function}
\label{sec:mckgf}

\YU{
A $2M \times 2M$ Keldysh Green function matrix ${\bm g}_{A}$ consists of $2 \times 2$ sub-matrices in the normal Keldysh space. 
A ($m,m'$) component ($m,m'=1,\cdots,M$) is, 
%------------------------------------------------------------------------------
\begin{widetext}
\begin{align}
\left[ {\bm g}_{A}^{ \{ \chi_j \} }(t,t') \right]_{m,m'} =&
\left[\begin{array}{cc}
g_{A}^{\{ \chi_j \} , m+,m'+} & g_{A}^{\{ \chi_j \} , m+,m'-} \\
g_{A}^{\{ \chi_j \} , m-,m'+} & g_{A}^{\{ \chi_j \} , m-,m'-}
\end{array}\right]
= -i e^{-i \epsilon_A (t-t')} \nonumber \\ &\times
\left \{
\begin{array}{cc}
e^{i \sum_{j=m'}^{m-1} \delta \chi_j s_A}
\left[ \begin{array}{cc}
f^{\bar{\chi}}_{A,m-m'}(\epsilon_A) & f^{\bar{\chi}}_{A,m-m'+1}(\epsilon_A) e^{-i {\bar{\chi}} s_A/M} \\
f^{\bar{\chi}}_{A,m-m'-1}(\epsilon_A) e^{i {\bar{\chi}} s_A/M} & f^{\bar{\chi}}_{A,m-m'}(\epsilon_A)
\end{array} \right]
& 
(m > m')
\\
\left[
\begin{array}{cc}
f^{\bar{\chi}}_{A,0}(\epsilon_A) \theta(t-t') - f^{\bar{\chi}}_{A,M}(\epsilon_A) \theta(t'-t) 
& f^{\bar{\chi}}_{A,1}(\epsilon_A) e^{-i {\bar{\chi}} s_A/M}
\\
- f^{\bar{\chi}}_{A,M-1}(\epsilon_A) e^{i {\bar{\chi}} s_A/M}
& f^{\bar{\chi}}_{A,0}(\epsilon_A) \theta(t'-t) - f^{\bar{\chi}}_{A,M}(\epsilon_A) \theta(t-t')
\end{array}
\right] & (m=m') \\
e^{-i \sum_{j=m}^{m'-1} \delta \chi_j s_A}
\left[
\begin{array}{cc}
-f^{\bar{\chi}}_{A,M+m-m'}(\epsilon_A) & -f^{\bar{\chi}}_{A,M+m-m'+1}(\epsilon_A) e^{-i {\bar{\chi}} s_A/M} \\
-f^{\bar{\chi}}_{A,M+m-m'-1}(\epsilon_A) e^{i {\bar{\chi}} s_A/M} & -f^{\bar{\chi}}_{A,M+m-m'}(\epsilon_A)
\end{array}
\right] & (m<m')
\end{array}
\right.
, 
\label{mmckgf_t}
\end{align}
\end{widetext}
%------------------------------------------------------------------------------
where we write $s_A=s_A(\epsilon_A)$. 
The modified Fermi distribution function is given by, 
%------------------------------------------------------------------------------
\begin{align}
f^{ \bar{\chi} }_{A, m}(\omega)=\frac{ e^{- m (1-i \bar{\chi}/M) s_A(\omega)} }{1+e^{- M (1-i \bar{\chi}/M) s_A(\omega)}} \, . 
\label{mfdf}
\end{align}
%------------------------------------------------------------------------------
Equation (\ref{mmckgf_t}) is Eq.~(57) in Ref.~\onlinecite{YU2017} replaced $\chi_j$ with $\chi_j s_A$. 
For detailed derivations, see Ref.~\onlinecite{YU2017}.

The $2M \times 2M$ Keldysh Green function matrix ${\bm g}_{A}^{ \bar{\chi} }$ in Eq.~(\ref{eqn:unimgf}) is obtained after the Fourier transform in time. 
It is a block skew-circulant; 
%------------------------------------------------------------------------------
\begin{align}
{\bm g}_{A}^{ \bar{\chi} }(\omega) = \left[ \begin{array}{ccccc}
{\bm A}_0 & -{\bm A}_{M-1} & -{\bm A}_{M-2} & \cdots & -{\bm A}_{1} \\
{\bm A}_1 &  {\bm A}_{0}   & -{\bm A}_{M-1} & \cdots & -{\bm A}_{2} \\
{\bm A}_2 &  {\bm A}_{1}   &  {\bm A}_{0} & \cdots & -{\bm A}_{3} \\
\vdots & \vdots & \vdots & \ddots & \vdots \\
{\bm A}_{M-1} & {\bm A}_{M-2} & {\bm A}_{M-3} & \cdots & {\bm A}_{0}
\end{array} \right] \, . \label{mmckgf}
\end{align}
%------------------------------------------------------------------------------
A diagonal component is, 
%------------------------------------------------------------------------------
\begin{align}
{\bm A}_0 =& {\rm P} \frac{1}{\omega - \epsilon_{A}} \, {\bm \tau}_3 -	2 \pi i \, \delta(\omega - \epsilon_{A}) 
\nonumber \\ & \times
\left[ \begin{array}{cc}
1/2-f_{A,M}^{\bar{\chi}}(\omega) & f_{A,1}^{\bar{\chi}} (\omega) e^{-i \frac{\bar{\chi}}{M} s_A(\omega)} \\
-f_{A,M-1}^{\bar{\chi}} (\omega) e^{i \frac{\bar{\chi}}{M} s_A(\omega)} & 1/2-f_{A,M}^{\bar{\chi}}(\omega)
\end{array} \right] \, 
\label{eqn:g_a_diag}
\end{align}
%------------------------------------------------------------------------------
where ${\bm \tau}_3={\rm diag} (1,-1) $. 
The phase factor $e^{-i \bar{\chi} s_A(\omega)/M}$ is equivalent to what appears in the full-counting statistics~\cite{Levitov1996,Nazarov2003,BUGS2006,Esposito2009} of heat current~\cite{KindermannPRB2004,SD2007,Heikkila2009,Laakso2010,Golubev2013,UEAKT2014,Wollfarth2014}. 
The delta function is 
%------------------------------------------------------------------------------
\begin{subequations}
\begin{align}
\delta(\omega)={\rm Im} \, \frac{1}{\pi(\omega - i \eta)}
\, ,
\label{eqn:deltaf}
\end{align}
%------------------------------------------------------------------------------
where $\eta$ is a positive infinitesimal. 
${\rm P}$ stands for the Cauchy principal value, 
%------------------------------------------------------------------------------
\begin{align}
{\rm P} \, \frac{1}{\omega}={\rm Re} \, \frac{1}{\omega - i \eta}. 
\label{eqn:cauchyp}
\end{align}
\end{subequations}
%------------------------------------------------------------------------------
An off-diagonal component is, 
%------------------------------------------------------------------------------
\begin{align}
{\bm A}_m =& - 2 \pi i \, \delta(\omega - \epsilon_{A}) 
\nonumber \\
& \times
\left[ \begin{array}{cc}
f_{A,m}^{\bar{\chi}}(\omega) & f_{A,m+1}^{\bar{\chi}}(\omega) e^{-i \frac{\bar{\chi}}{M} s_A(\omega)} \\
f_{A,m-1}^{\bar{\chi}}(\omega) e^{i \frac{\bar{\chi}}{M} s_A(\omega)} & f_{A,m}^{\bar{\chi}}(\omega)
\end{array} \right]
\, , 
\label{eqn:g_a_off_diag}
\end{align}
%------------------------------------------------------------------------------
where $m=1,\cdots,M-1$. 

The block-skew circulant matrix is block-diagonalized by the following discrete Fourier transform~\cite{YU2015,YU2017}, 
%------------------------------------------------------------------------------
\begin{align}
{\mathbf g}_{A}^{\lambda_\ell-\bar{\chi} s_A(\omega) /M} = \sum_{m-m'=0}^{M-1} \left[{\bm g}_{A}^{\bar{\chi}} \right]_{m,m'} e^{i \pi \frac{2 \ell+1}{M} (m-m')}
\label{dft}
\, ,  
\end{align}
%------------------------------------------------------------------------------
where $\lambda_\ell = \pi [ 1- (2 \ell+1)/M ]$ and the $2 \times 2$ Green function matrix in the LHS is, 
%------------------------------------------------------------------------------
\begin{align}
{\mathbf g}_{A}^{\lambda}(\omega) =& {\rm P} \frac{1}{\omega-\epsilon_A} {\bm \tau}_3 -2 \pi i \delta(\omega-\epsilon_A)
\nonumber \\
& \times
\left[ \begin{array}{cc} 1/2-f_{A,\lambda}^+(\omega) & f_{A,\lambda}^+(\omega) e^{i \lambda} \\ -f_{A,\lambda}^-(\omega) e^{-i \lambda} & 1/2-f_{A,\lambda}^+(\omega) \end{array} \right]
\, .
\label{eqn:free_Keldysh_gf}
\end{align}
%------------------------------------------------------------------------------
Equation (\ref{eqn:free_Keldysh_gf}) is the modified Keldysh Green function appeared in the theory of the full-counting statistics~\cite{UGS2006,BUGS2006,Gogolin2006,SU2008,US2009,Esposito2009,Urban2010,Sakano2011,Novotny2011,UEUA2013,Tang2014,SoeaneSouto2015}. 
Precisely, the standard scheme of the full-counting statistics is based on the two-time measurement protocol, which means that the measurement is done twice: once in the beginning and once in the end~\cite{SU2008,Esposito2009}. 
In the present case, the measurement is effectively done once in the end~\cite{YU2018fqmt}. 
Because of this difference, the electron and hole distribution functions are modified; 
%------------------------------------------------------------------------------
\begin{align}
f_{A,\lambda}^+(\omega) = \frac{f_{A}^+(\omega)}{f_{A}^+(\omega) +f_{A}^-(\omega) e^{i \lambda}}
\, , \;\;\;\;
f_{A,\lambda}^-(\omega) = 1-f_{A,\lambda}^+(\omega)
\, . 
\end{align}
%------------------------------------------------------------------------------

The Fourier transform of the $2M \times 2M$ Keldysh Green function matrix for subsystem $B$, see Eq.~(\ref{gfB}), is, 
%------------------------------------------------------------------------------
\begin{align}
{\bm g}_{B}(\omega) =& {\bm 1} \otimes {\mathbf g}_B(\omega) \, ,
\label{eqn:mmckgfB}
\\
{\mathbf g}_B(\omega) =& {\rm P} \frac{1}{\omega - \epsilon_{B}} \, {\bm \tau}_3 - 2 \pi i \, \delta(\omega - \epsilon_{B}) 
\nonumber \\
& \times
\left[ \begin{array}{cc} 1/2-f_B^+ (\omega) & f_B^+ (\omega) \\ -f_B^- (\omega) & 1/2-f_B^+ (\omega) \end{array} \right]
\, , 
\label{eqn:mmckgfB_omega}
\end{align}
%------------------------------------------------------------------------------
where ${\bm 1}$ is the $M \times M$ identity matrix. 

}

\section{Dot Green function matrix}
\label{sec:dotgfm}

The self-energy of the Dyson equation (\ref{eqn:dyson_22_keldysh}) is $\sum_k \left( J_L^2 {\mathbf g}_{Lk}^\lambda + J_R^2 {\mathbf g}_{Rk} \right) = {\mathbf \Sigma}_L^\lambda + {\mathbf \Sigma}_R^{\lambda=0}$ where, 
%------------------------------------------------------------------------------
\begin{align}
{\mathbf \Sigma}_r^\lambda(\omega)
=
-i 
\frac{\Gamma_r}{2}
\left[
\begin{array}{cc}
1-2 f_{r, \lambda}^+(\omega) & 2 f_{r, \lambda}^+(\omega) e^{i \lambda} \\
- 2 f_{r, \lambda}^-(\omega) e^{-i \lambda} & 1-2 f_{r, \lambda}^+(\omega)
\end{array}
\right]
\, .
\end{align}
%------------------------------------------------------------------------------
\YU{
By paying attention to Eqs.~(\ref{eqn:deltaf}) and (\ref{eqn:cauchyp}), the matrix inverse of the bare dot Green function matrix, i.e. Eq.~(\ref{eqn:mmckgfB_omega}) replaced $B$ with $D$, is calculated as, 
%------------------------------------------------------------------------------
%\begin{widetext}
\begin{align}
{\mathbf g}_D(\omega)^{-1} =& (\omega-\epsilon_D) {\bm \tau}_3
\nonumber \\
&+ 2 i \eta {\bm \tau}_3 \left[ \begin{array}{cc} 1/2-f_D^+(\omega) & f_D^+(\omega)  \\ -f_D^-(\omega) & 1/2-f_D^+(\omega) \end{array} \right] {\bm \tau}_3 \, . 
\label{eqn:invgd}
\end{align}
%------------------------------------------------------------------------------
The second line of the RHS depends on the parameters $\beta_D$ and $\mu_D$ characterizing the initial dot state, through the electron distribution function $f_D^+(\omega)=1/(e^{-\beta_D(\omega - \mu_D)}+1)$. 
It is noticed that these parameters disappear in the steady state, as we anticipated, because the second line of Eq.~(\ref{eqn:invgd}) is proportional to the positive infinitesimal $\eta$ and thus is negligible as compared with the self-energy in the Dyson equation (\ref{eqn:dyson_22_keldysh}). 
Then the solution is independent of these parameters as, 
}
%------------------------------------------------------------------------------
\begin{widetext}
\begin{align}
{\mathcal G}_{D}^\lambda(\omega)
=&
\frac{ {\mathcal G}_{D}(\omega) }{\Omega_{1,\lambda}(\omega)}
+
\rho(\omega)
\frac{\Gamma_L}{\Gamma}
\frac{2 \pi i (1-e^{i \lambda})}
{\tilde{f}_L^+(\omega) + \tilde{f}_L^-(\omega) e^{i \lambda}}
%\nonumber \\ & \times
\left[
\begin{array}{cc}
f_L^+(\omega) f_L^-(\omega) & f_L^+(\omega)^2  \\ 
f_L^-(\omega)^2 & f_L^+(\omega) f_L^-(\omega)
\end{array}
\right]
\, , 
\label{hgf}
\end{align}
%------------------------------------------------------------------------------
where the DOS of dot is $\rho(\omega) = {\mathcal T}(\omega) \Gamma/(2\pi \Gamma_L \Gamma_R)$ and, 
%------------------------------------------------------------------------------
\begin{align}
{\mathcal G}_{D}(\omega)
&=
\frac{2 \pi}{\Gamma}
\rho(\omega)
\left[
\begin{array}{cc}
\omega-\epsilon_D - i \sum_r \Gamma_r [1/2 - f_{r}^+(\omega)] & 
-i \sum_r \Gamma_r f_{r}^+(\omega) \\
 i \sum_r \Gamma_r f_{r}^-(\omega) & 
\epsilon_D-\omega - i \sum_r \Gamma_r [1/2 - f_{r}^+(\omega)]
\end{array}
\right]
\, . 
\end{align}
%------------------------------------------------------------------------------
The following relations can be derived by exploiting Eq.~(\ref{eqn:continteg}); 
%------------------------------------------------------------------------------
\begin{align}
\sum_{\ell=0}^{M-1} \frac{1}{\Omega_{1,\lambda_\ell-\chi s_A(\omega)/M}(\omega)}
&= M - \frac{\partial_{\epsilon_D} \ln \Omega_{M,-\chi s_A(\omega)/M}(\omega)}{\partial_{\epsilon_D} \ln \rho(\omega)} \, ,
\\
\sum_{\ell=0}^{M-1} \frac{1-e^{i \lambda_\ell -i \chi s_A(\omega)/M}} {\tilde{f}_L^+(\omega)+\tilde{f}_L^-(\omega) e^{i \lambda_\ell -i \chi s_A(\omega)/M}}
&= \frac{ \partial_{\epsilon_D} \ln \Omega_{M,-\chi s_A(\omega)} (\omega) } {\partial_{\epsilon_D} \ln \rho(\omega){\mathcal T}(\omega)(f_R^+(\omega) - f_L^+(\omega))} \, . 
\end{align}
%------------------------------------------------------------------------------
Then the local Green function in the replicated Keldysh space is, 
%------------------------------------------------------------------------------
\begin{align}
\left[{\bm G}_{D}^\chi(\omega) \right]_{m,m}
&=
\frac{1}{M}
\sum_{\ell=0}^{M-1}
{\mathcal G}_{D}^{\lambda_\ell-\chi s_A(\omega)/M}(\omega)
=
{\mathcal G}_{D}(\omega)
\\
&-
\frac{1}{M} 
\frac{ \partial_{\epsilon_D} \ln \Omega_{M,-\chi s_A(\omega)}(\omega) }{\partial_{\epsilon_D} \ln \rho(\omega)}
\biggl(
{\mathcal G}_{D}(\omega)
-
%\nonumber \\ & \times
\frac{2 \pi i (\Gamma_L/\Gamma) \rho(\omega)}
{ {\mathcal T} (\omega) (f_R^+(\omega) - f_L^+(\omega) )}
\left[
\begin{array}{cc}
f_L^+(\omega) f_L^-(\omega) & f_L^+(\omega)^2  \\ 
f_L^-(\omega)^2 & f_L^+(\omega) f_L^-(\omega)
\end{array}
\right]
\biggl)
\, . 
\label{eqn:rlgfm}
\end{align}
%------------------------------------------------------------------------------
The result does not change when we account for the spin degree of freedom.

\end{widetext}

\section{Summation}
\label{sec:summation}

\begin{figure}[hb]
\includegraphics[width=0.6 \columnwidth]{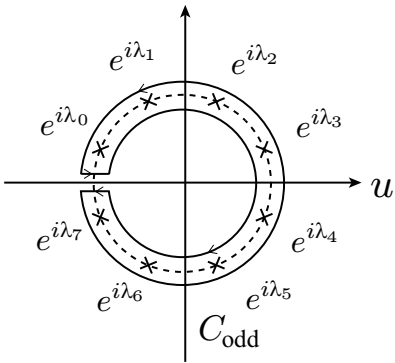}
\caption{Contour $C_{\rm odd}$ enclosing poles $e^{i \lambda_\ell}$ ($\ell=0,\cdots, M-1$) [$M=8$ in this panel]. 
The dotted line indicates a unit circle. 
}
\label{cont_integ}
\end{figure}

The summation over $\ell$ in Eq.~(\ref{maires}) can be done by exploiting the following relation~\cite{Casini2009}. 
Let $g$ be a function. 
The summation is rewritten as the contour integral as, 
%------------------------------------------------------------------------------
\begin{align}
\sum_{\ell=0}^{M-1}
g(e^{i \lambda_\ell}) &= \int_{C_{\rm odd}} \frac{d u}{2 \pi i} \sum_{\ell=0}^{M-1} \frac{g(u)}{u-e^{i \lambda_\ell}}
\nonumber \\
&=\int_{C_{\rm odd}} \frac{d u}{2 \pi i} \frac{-M(-u)^{M-1}}{1+(-u)^M} g(u)
\, ,
\label{eqn:continteg}
\end{align}
%------------------------------------------------------------------------------
where $\lambda_\ell=\pi [1-(2 \ell+1)/M]$. 
The contour $C_{\rm odd}$ encloses $M$ poles $e^{i \lambda_\ell}$ ($\ell=0,\cdots, M-1$), see Fig.~\ref{cont_integ}.

%----------------------------------------------------------
%----------------------------------------------------------


\begin{thebibliography}{99}
\bibitem{note1}
Precisely, the R\'enyi entropy of probability distribution $P=(p_1, \cdots, p_n)$ is~\cite{Renyi1960}, 
%------------------------------------------------------------------------------
$\log \left ( \sum_{k=1}^n p_k^{\alpha} \right )/(\alpha-1)$
%------------------------------------------------------------------------------
where $\alpha \neq 1$. 
In the present paper, we call $\sum_{k=1}^n p_k^{\alpha}$ as the R\'enyi entropy
\end{thebibliography}
\end{document}